\documentclass[journal=jacsat,manuscript=article]{achemso}

\usepackage{chemformula} 
\usepackage[T1]{fontenc} 

\author{Aurelian Loirette-Pelous}
\altaffiliation{Current address: Materials Physics Center, CSIC-UPV/EHU, Donostia-San Sebastián, Spain
}
\email{aurelianjoan.loirettepelous@ehu.eus}

\author{Jean-Jacques Greffet}
\email{jean-jacques.greffet@institutoptique.fr}

\affiliation[Unknown University]{Université Paris-Saclay, Institut d'Optique Graduate School, CNRS, Laboratoire Charles Fabry}

\title[An \textsf{achemso} demo]
  {Theory of photoluminescence by metallic structures}

\abbreviations{IR,NMR,UV}

\begin{document}







\begin{abstract}
Light emission by metals at room temperature is quenched by fast relaxation processes. Nevertheless, Mooradian reported in 1969 the observation of photoluminescence by metals pumped by a laser. Strikingly, while it is currently at the heart of many promising applications, this phenomenon is still poorly understood. In this work, we report a theory which reproduces quantitatively previously published experimental data. We first provide a general formula that relates the emitted power for a frequency, direction and polarization state to a sum over all transitions involving matrix elements, electronic distribution of all bands and the Green tensor. We then consider the case of intraband recombination and derive a closed-form expression of the emitted power depending only on macroscopic quantities. This formula, which is a generalization of Kirchhoff's law, answers many of the open questions related to intraband photoluminescence.
\end{abstract}


\section{Introduction}

Photoluminescence, broadly defined as the process of light emission by a material following the absorption of an incident radiation, is a major process for light generation in physics. It is easily observable in materials with a gap such as semiconductors or dye molecules. Light emission is then due to electron-hole recombination and called fluorescence. Incandescence is an alternative light emission mechanism, that arises due to heating of the emitting material under strong illumination. An incident radiation can also produce Raman scattering,  which is an ultrafast coherent process that does not involve electronic population modification. 

Metals are known to emit light by incandescence but fluorescence is usually quenched. Indeed, electron-electron and electron-phonon interactions provides a fast relaxation path for excited electrons. Owing to the absence of a gap, nothing prevents a full relaxation so that no light emission is observed. However, observation of photoluminescence in the visible and near infrared was reported by Mooradian in 1969 \cite{mooradian1969}. This phenomenon was surprising given that metal heating was negligible in the experimental conditions. This effect is characterized by a very low quantum yield on the order of $\sim 10^{-10}$ \cite{mooradian1969}, that has hindered its study for decades. {Nevertheless,} observation of luminescence by metals is {nowadays} a simple task thanks to recent advances in the fabrication and characterization of single metallic nanoparticles \cite{beversluis2003,gaiduk2011correlated,link2011}. Due to a strong plasmonic resonance, quantum yields can be enhanced up to $\sim 10^{-5}-10^{-6}$ \cite{yorulmaz2012luminescence}, so that photoluminescence is now a source of detrimental background signal when performing Raman scattering measurements \cite{baumberg2015}. By contrast, photoluminescence from metallic nanoparticles has become an useful tool. For example, it enables to create new luminescent markers in biomedecine \cite{wu2010high,carattino2016background}, to supervise chemical reactions \cite{zheng2014single,zheng2015plasmon,hill2015single} or to measure the temperature at the nanoscale \cite{cahill2014,cahill2016,carattino2018gold,jones2018photothermal,link2019anti,sheldon2019,barella2020,baffou2021}, among other applications.

Despite a large number of studies and these practical applications, the physical origin of photoluminescence by metallic nanostructures is still under debate \cite{baffou2021,cai2021light}. On the one hand, the long standing explanation in term of interband transitions \cite{mooradian1969,boyd1986photoinduced,apell1988photoluminescence,tagliabue2023} is still widely accepted for emission at frequencies beyond the interband threshold. On the other hand, many evidences of emission at frequencies lower than the interband threshold have been observed in the past twenty years \cite{beversluis2003,eustis2005aspect,link2011,cahill2014,baumberg2015,ren2016,baumberg2017,link2018}, revealing the necessity to consider other mechanisms. Therefore, current alternative theories focus on processes occurring within the metal conduction band. Intraband recombinations have been first proposed as the origin of the luminescence \cite{beversluis2003}, later followed by other works \cite{ren2016,lupton2015,link2019anti,link2020increased}. However, the peculiar temperature dependence of the anti-Stokes spectrum used for thermometry was first explained by a mechanism based on electronic Raman scattering \cite{cahill2014,baumberg2015,baumberg2017}. Nevertheless, later theoretical work showed that the temperature dependence of the anti-Stokes spectrum can also been explained with an intraband recombination process, considering that a non-equilibrium electronic distribution is induced by the pumping \cite{sivan2021theory}. Finally, let us note that this theory has predicted a peculiar scaling of the emitted power with the pumping strength \cite{sivan2021theory}. Interestingly, recent experiments have observed this scaling  \cite{sivan2023crossover}, giving support to the hypothesis of an intraband recombination mechanism.

Leaving aside the controversy on the physical mechanism involved in photoluminescence by metals, there is still a lack of models able to predict quantitatively the emitted power. Such a model must account for two effects: a nonequilibrium electronic distribution and the emission in a metallic environment. The first theory enabling comparisons with experiments has been proposed in Ref. \citenum{link2018}, considering a non-equilibrium hot carriers distribution and including emission from both interband and intraband electron-hole recombinations. A good agreement with experimental measurements was obtained for the Stokes part of the spectrum but the anti-Stokes part of the spectrum could not be modeled due the assumption of zero electronic temperature. This issue has been tackled in the case of intraband recombinations in Ref. \citenum{sivan2021theory}. Yet, in both theories the metallic environment is included by inserting a photonic density of states. To date, the most elaborated theory has been proposed in Ref. \citenum{tagliabue2023} to model the PL of planar gold flakes. It features a calculation of the nonequilibrium electronic distribution for both intraband and interband transitions, combined with a calculation of the corresponding transition matrix elements and the calculation of the Green tensor. Yet, while a good agreement is found with experimental data for emission from interband transitions, it fails to reproduce the emission from intraband transitions. This indicates that the intraband matrix elements are not correctly taken into account or that an alternative mechanism should be introduced.

Here, we introduce a quantitative theory of photoluminescence by metals that covers both intraband and interband transitions. In the particular case of intraband photoluminescence, we circumvent the calculation of matrix elements by deriving a closed-form expression which depends on the absorption cross-section. To this end, we use a statistical physics approach of light emission. The emitted power is computed from the knowledge of fluctuating currents which are included in Maxwell equations in the spirit of a Langevin model. We derive a fluctuation relation connecting the current density correlation function with the non-equilibrium electronic distribution. This formulation is exact and general but difficult to implement. In the case of intraband photoluminescence, we show that it can be drastically simplified by inserting further approximations. We manage to cast the result in the form of a generalized Kirchhoff's law so that the emitted power can be expressed as the product of the absorption cross section of the metallic nanostructure and a generalized blackbody radiance. When using non-equilibrium conduction band distributions evaluated by solving the Boltzmann equation under pumping by a continuous wave (CW) laser, we obtain analytical formulas that enable to solve many debates related to intraband CW photoluminescence. Finally, we proceed to quantitative comparisons with experimental data.

\section{Results and discussion}

\subsection{General theory}

modeling light emission by a metallic nanostructure is a subtle issue. On the one hand, a quantum point of view is implicitly taken when referring to electron-hole recombinations. On the other hand, radiation problems associated to emitters embedded in a specific (possibly resonant) electromagnetic environment are naturally tackled in the classical framework of Maxwell equations, with electric dipoles as source terms. Hence, a quantitative theory of photoluminescence by metals must connect the classical and quantum points of view on light emission. In this section, we bridge this gap using a statistical physics theory of light emission \cite{rytov1989,benisty2022introduction}. The cornerstone of this framework is the cross spectral density $W_{j^{\dagger}j} (\boldsymbol{r}_1, \boldsymbol{r}_2, \omega)$, that quantifies the correlation of the fluctuating current density operators $\hat{\boldsymbol{j} }$ at positions $\boldsymbol{r}_1$ and $\boldsymbol{r}_2$ and at frequency $\omega$. We will show how this quantity is connected to the two points of view underlying the emitted light. This will eventually provide a general relation connecting the emitted power with the non-equilibrium electronic distribution.

First, let us remark that $W_{j^{\dagger}j}$ is the source term of the classical radiation problem involved in the light emission. Indeed, Maxwell equations connect the radiated field $\hat{\boldsymbol{E}}$ and the current density $\hat{\boldsymbol{j} }$, that can be written using the Green tensor $\boldsymbol{G}$ as $\hat{\boldsymbol{E}}(\boldsymbol{r})~=~i \omega \mu_0 \int \boldsymbol{G}(\boldsymbol{r},\boldsymbol{r}',\omega) \hat{\boldsymbol{j}}(\boldsymbol{r}',\omega) d\boldsymbol{r}'$. We note that the Green tensor accounts for the electromagnetic environment including field enhancement and resonances. Hence, we show in Section A. of the Supporting Information that the power $dP_e$ emitted at a frequency $\omega$, in a solid angle $d\Omega_r$ centered around a position $\boldsymbol{r}$ in the far-field of a metallic structure and in a polarisation state $p$ is related to $W_{j^{\dagger}j}$ through the equation:

\begin{equation}\label{eq:pui_em_psd_abs_main}
    \frac{dP_e}{d\Omega_r}(\omega,\boldsymbol{u}_r,p) =  \sigma_{abs}(\omega,-\boldsymbol{u}_r, p)  \frac{ \omega^2}{8 \pi^3 c^2 }   \frac{ W_{j^{\dagger}j} (\omega) }{ 2 \omega \epsilon_0 \text{Im}[\epsilon(\omega)] },
\end{equation}

where $\boldsymbol{u}_r = \boldsymbol{r}/|\boldsymbol{r}|$ is the direction of emission defined by the position $\boldsymbol{r}$, and $c$ is the vacuum speed of light. $\sigma_{abs}^{}(\omega,-\boldsymbol{u}_r,p^*)$ is the absorption cross section of the metallic structure, calculated for an incident plane wave of same frequency but reverse direction and polarization handedness compared to the far field plane wave emitted by photoluminescence. This quantity contains the information on the Green tensor, as shown in Section A. of the Supporting Information. Let us also point out that the presence of the absorption cross section in a calculation of light emission is a consequence of Lorentz reciprocity \cite{Collin60}. The dimensionless relative permittivity of the emitting metal is noted $\epsilon(\omega,\boldsymbol{r})$. Finally, it is assumed that the metal is isotropic and homogeneous, so that $W_{j^{\dagger}j}$ and $\epsilon$ are scalars and their spatial dependence can be dropped.

Second, $W_{j^{\dagger}j}$ describes the statistical behavior of the fluctuating current densities in the metal. In Section B.1. of the Supporting Information, we show that for a homogeneous and isotropic electron gas with arbitrary mean occupation number of the electronic states, the cross spectral density of the current density can be cast into the form:
\begin{equation}\label{eq:densite_spectrale_puissance_main}
       W_{j^{\dagger}j} (\omega) = \frac{\epsilon_0 \omega_p^2}{m} \frac{2\pi}{N} \sum_{n,n'=1}^{N}  |p^{}_{n,n'}|^2  f_{n'} [1 - f_{n}] \delta( \omega - [\omega_{n'} - \omega_{n}] ),
\end{equation}

where $\epsilon_0$ is the vacuum permittivity, $\omega_p$ is the metal plasma frequency, $m$ is the electron mass, $N$ is the total electron number in the metal, $n=1,2,...,N$ labels the electronic states without specifying a band, $\hbar \omega_n$ labels their corresponding energy and $f_{n}$ their mean occupation number, while $p^{}_{n,n'}$ is the transition matrix element between state $n$ and state $n'$. Importantly, let us emphasize that Eq. (\ref{eq:densite_spectrale_puissance_main}) describes the current density fluctuations in any nonequilibrium condition, i.e. the $f_n$ can take any value. Remarkably, in the specific case of thermodynamic equilibrium, the mean occupation number follows the Fermi-Dirac distribution, that is $f_n= 1/[ \exp(\frac{\hbar \omega_n - \mu_F}{k_B T}) + 1]$ where $k_B$ is the Boltzmann constant, $T$ is the temperature of the equilibrium electron gas and $\mu_F$ its Fermi level. Eq. (\ref{eq:densite_spectrale_puissance_main}) then reduces to (see Sections B.2. and B.3. of the Supporting Information):
\begin{equation}\label{eq:FDT_Wjj}
    W_{j^{\dagger}j,eq}^{} (\omega) = 2 \omega \epsilon_0 \text{Im}[\epsilon_{eq}(\omega)] \frac{\hbar \omega}{ \exp \big( \frac{\hbar \omega}{k_B T} \big) - 1},
\end{equation}

where $\epsilon_{eq}$ is the metal relative permittivity at equilibrium. Equation (\ref{eq:FDT_Wjj}) is known as the Fluctuation-Dissipation Theorem (FDT) \cite{callen1951irreversibility,kubo1966fluctuation}. Indeed, it connects the current fluctuations to the imaginary part of the permittivity, that quantifies how the electrons dissipate the energy provided by an external radiation. Hence, we note that the fluctuation relation Eq. (\ref{eq:densite_spectrale_puissance_main}) generalizes the FDT Eq. (\ref{eq:FDT_Wjj}) to nonequilibrium situations. Yet, we also note that it cannot be related, in general, to $\text{Im}[\epsilon_{}^{}]$ so that it is not a fluctuation-dissipation relation.

In summary, the fluctuation relation Eq. (\ref{eq:densite_spectrale_puissance_main}) enables to establish the expected link between the quantum matter and classical radiation pictures of light emission: On the one hand, $W_{j^{\dagger}j}$ is the source term of the classical radiation problem involved in the light emission, as shown in Eq. (\ref{eq:pui_em_psd_abs_main}). On the other hand, the right hand side of Eq. (\ref{eq:densite_spectrale_puissance_main}) clearly displays the quantum point of view through the product $f_{n'}[ 1-  f_{n} ] \delta( \omega - [\omega_{n'} - \omega_{n}] )$, that corresponds to the recombination probability of an electron on a state with energy $\hbar \omega_{n} + \hbar \omega$ with a hole on a state with energy $\hbar \omega_{n}$.

Lastly, we insert Eq. (\ref{eq:densite_spectrale_puissance_main}) into Eq. (\ref{eq:pui_em_psd_abs_main}), which shows that the photoluminescence power can be cast into the form: 

\begin{equation}\label{eq:pui_ray_he}
     \frac{dP_e}{d\Omega_r} (\omega,\boldsymbol{u}_r,p)
     =  \sigma_{abs}^{} (\omega,-\boldsymbol{u}_r,p^*)   \frac{\omega^2}{8 \pi^3 c^2} \frac{ \omega_{p}^2 }{2 \omega \text{Im}[\epsilon_{}(\omega)] } \frac{2\pi}{N} \sum_{n,n'=1}^{N}  |p^{}_{n,n'}|^2  f_{n'} [1 - f_{n}] \delta( \omega - [\omega_{n'} - \omega_{n}] ).
\end{equation}

Eq. (\ref{eq:pui_ray_he}) is the first main result of this work. It provides a framework to calculate quantitatively the emitted power at each frequency, direction and polarization state by a metallic structure. Remarkably, Eq. (\ref{eq:pui_ray_he}) is of very broad generality: (i) the absorption cross section is not the absorption by a volume of the bare metal but the absorption of a system. It can be, e.g. the absorption by a metallic nanoparticle in vacuum, or the same metallic nanoparticle placed in a cavity. Hence, it takes into account the local photonic density of states in the metal. (ii) Eq. (\ref{eq:pui_ray_he}) was derived for a stationary excitation. Yet, it can also be used for incident pulses under a quasistationary approximation using a time-dependent electronic population and permittivity. (iii) Similarly, metal inhomogeneities can be accounted for using a position-dependent absorption cross section density \cite{Greffet2018}, electronic population and permittivity. (iv) Light emission can be calculated with Eq. (\ref{eq:pui_ray_he}) for nonequilibrium electrons. Hence, it entails light emission by metallic bodies possibly driven out-of-equilibrium by an external (pulsed or continuous) pumping. (v) Eq. (\ref{eq:pui_ray_he}) encompasses both intraband and interband photoluminescence, as no hypothesis on the type of transitions has been made at this point.

\subsection{Generalized Kirchhoff's law for emission from intraband transitions}

Modeling photoluminescence by metallic structures with Eq. (\ref{eq:pui_ray_he}) requires the knowledge of the band structure and of the transition matrix elements of the metal. While the derivation of these microscopic quantities has been performed in several publications \cite{sundararaman2014theoretical,bernardi2015theory,atwater2016nonradiative,coudert2022unified,tagliabue2023}, it demands complex and cumbersome numerical investigations. In this section, we specifically focus on emission arising from intraband transitions and introduce an approximation that circumvent this difficulty, enabling to compute the emitted power knowing the absorption cross section. 

We start by assuming that the density of states of the conduction electrons, as well as the transition matrix elements corresponding to intraband transitions are constant over the electronic states, and equal to their value at the metal Fermi level. Let us note that this approximation is usually justified in calculations of the absorption by metals in the mid/far IR range \cite{ashcroft2022solid,ziman2001electrons}, but requires more investigations in the visible/near-IR range that are out of the scope of this work. Here, we will instead evaluate the impact of the approximation in the next section through comparison with experiments. We also assume that the mean occupation numbers $f_n$ of the conduction band electronic states follow a well defined distribution $f(E)$ that only depends on energy $E$. These hypotheses eventually enable to cast the fluctuation relation Eq. (\ref{eq:densite_spectrale_puissance_main}) into the form (see Section C. of the Supporting Information for a detailed derivation):

\begin{equation}\label{eq:fluctuation_dissipation_theorem_generalized_main}
     W_{j^{\dagger}j}^{intra} (\omega)= 2\epsilon_0 \omega \text{Im}[\epsilon_{eq}^{intra}(\omega)] \times \Theta^{intra}(\omega),
\end{equation}

where $\epsilon_{eq}^{intra}$ is the relative permittivity of the metal at equilibrium and limited to intraband transitions, and where:

\begin{equation}\label{eq:def_teta_gen}
    \Theta^{intra}(\omega) = \int f(E + \hbar \omega) [ 1-  f(E) ] dE,
\end{equation}

corresponds to the integral of the electron-hole recombination probabilities associated to indirect transitions of frequency $\omega$ in the conduction band. Let us emphasize that $f(E)$ can be any arbitrary distribution. Hence, Eqs.(\ref{eq:fluctuation_dissipation_theorem_generalized_main}),(\ref{eq:def_teta_gen}) appears as a generalization of the Fluctuation-Dissipation theorem Eq. (\ref{eq:FDT_Wjj}) to out-of-equilibrium statistics of conduction band electrons.

Finally, as long as nonequilibrium effects yield a small deviation of the imaginary part of the metal permittivity from the equilibrium one, inserting Eq. (\ref{eq:fluctuation_dissipation_theorem_generalized_main}) into Eq. (\ref{eq:pui_em_psd_abs_main}) enables to cast the expression of the emitted power into the form:

\begin{equation}\label{eq:kirchhoff_generalized}
     \frac{dP_e}{d\Omega_r}(\omega,\boldsymbol{u}_r,p)
     = \sigma_{abs}^{intra}(\omega,-\boldsymbol{u}_r,p^*) \frac{\omega^2}{8\pi^3 c^2} \Theta^{intra}(\omega),
\end{equation}

where $\sigma_{abs}^{intra}= (\text{Im}[\epsilon_{eq}^{intra}]/\text{Im}[\epsilon_{eq}])\times\sigma_{abs}^{}$ is the part of the total absorption that is due to intraband transitions. 

Eqs. (\ref{eq:fluctuation_dissipation_theorem_generalized_main}),(\ref{eq:kirchhoff_generalized}) constitute the second main result of this work. We now discuss further their properties.
Firstly, it is remarkable that at thermodynamic equilibrium, Eq. (\ref{eq:fluctuation_dissipation_theorem_generalized_main}) coincides with the equilibrium Fluctuation-Dissipation theorem Eq. (\ref{eq:FDT_Wjj}), and $\omega^2/(8\pi^3 c^2)\times \Theta^{intra}(\omega)$ yields the Planck's law. Eq. (\ref{eq:kirchhoff_generalized}) then reproduces the so-called Kirchhoff's law \cite{kirchhoff1860relation,Greffet2018}, that is the basic tool to model thermal radiation. Hence, Eqs. (\ref{eq:kirchhoff_generalized}), (\ref{eq:def_teta_gen}) identifies as a generalization of Kirchhoff's law to metals featuring an arbitrary conduction band electron distribution.

Secondly, as for the equilibrium Kirchhoff's law, let us stress that our approach is based on statistical physics. Hence, the detailed knowledge about the microscopic transition processes (phonon-assisted transitions, electron-assisted transitions etc...), as well as their respective weighting in the emission is not necessary. Instead, it is fully taken into account within macroscopic quantities: (i) the permittivity of the emitting metal, that underlies the absorption cross section; (ii) the distribution of the electrons in the conduction band.

Thirdly, as it only requires the knowledge of macroscopic quantities, the generalized Kirchhoff's law Eqs. (\ref{eq:kirchhoff_generalized}),(\ref{eq:def_teta_gen}) turns out to be way simpler to use than Eq. (\ref{eq:pui_ray_he}): (i) The electronic distributions can be routinely calculated with satisfying accuracy solving the Boltzmann equation in the relaxation time approximation \cite{ziman2001electrons,lundstrom2002fundamentals,lugovskoy1999ultrafast,dubi2019hot}. (ii) The absorption cross sections $\sigma_{abs}$ of the metallic nanostructure can be computed with Maxwell's solvers.
Further, the absorption cross section contains all the information on the plasmon resonance that shapes the emission spectrum, directivity and polarization; the electronic distribution contains all the information on the excitation frequency and, if defined, on the temperature.

\subsection{Application to emission arising from intraband transitions in the CW pumping regime}

\subsubsection{Generalized Kirchhoff's law under CW pumping}

We now focus on the intraband photoluminescence of single metallic nanoparticules illuminated by a CW pumping laser. This pumping scheme has been experimentally studied by many groups, so that it represents a convenient testbed for our generalized Kirchhoff's law. In particular, these efforts have resulted in an empirical guess for the emitted power \cite{baffou2021}:
\begin{equation}\label{eq:spectre_empirique}
    \frac{dP_e}{d\Omega_r}(\omega) \propto S_{LPR} (\omega) F(\hbar \omega - \hbar \omega_L,T) I_L,
\end{equation}

where $S_{LPR}$ is the localized plasmon resonance lineshape, $F(\hbar \omega,T)$ is a function of the frequency and temperature, $\omega_L$ is the frequency of the pump laser and $I_L$ its intensity at the sample position. However, several aspects of this empirical expression are still under debate. First, we note that $S_{LPR}$ has been modeled so far by the scattering cross section \cite{link2019anti} or by a quantity related to the local density of photonic states \cite{link2018}. Instead, Ref. \citenum{baffou2021} suggested that the absorption cross section should be rather used. Similarly, $F(\hbar \omega,T)$ is the Bose-Einstein distribution for some authors \cite{cahill2014,carattino2018gold,sheldon2019,barella2020} and the Fermi-Dirac distribution reduced to its Boltzmann component for others \cite{he2015surface,link2019anti}. Accordingly, $T$ has been proposed to be the temperature of the metal lattice (that is of phonons) in the former case and that of the electron gas in the latter. In this section, we compare Kirchhoff's law Eqs. (\ref{eq:kirchhoff_generalized}),(\ref{eq:def_teta_gen}) to Eq. (\ref{eq:spectre_empirique}) and address the open questions discussed above.

\begin{figure}[htbp]
    \centering
    \includegraphics[width=\columnwidth]{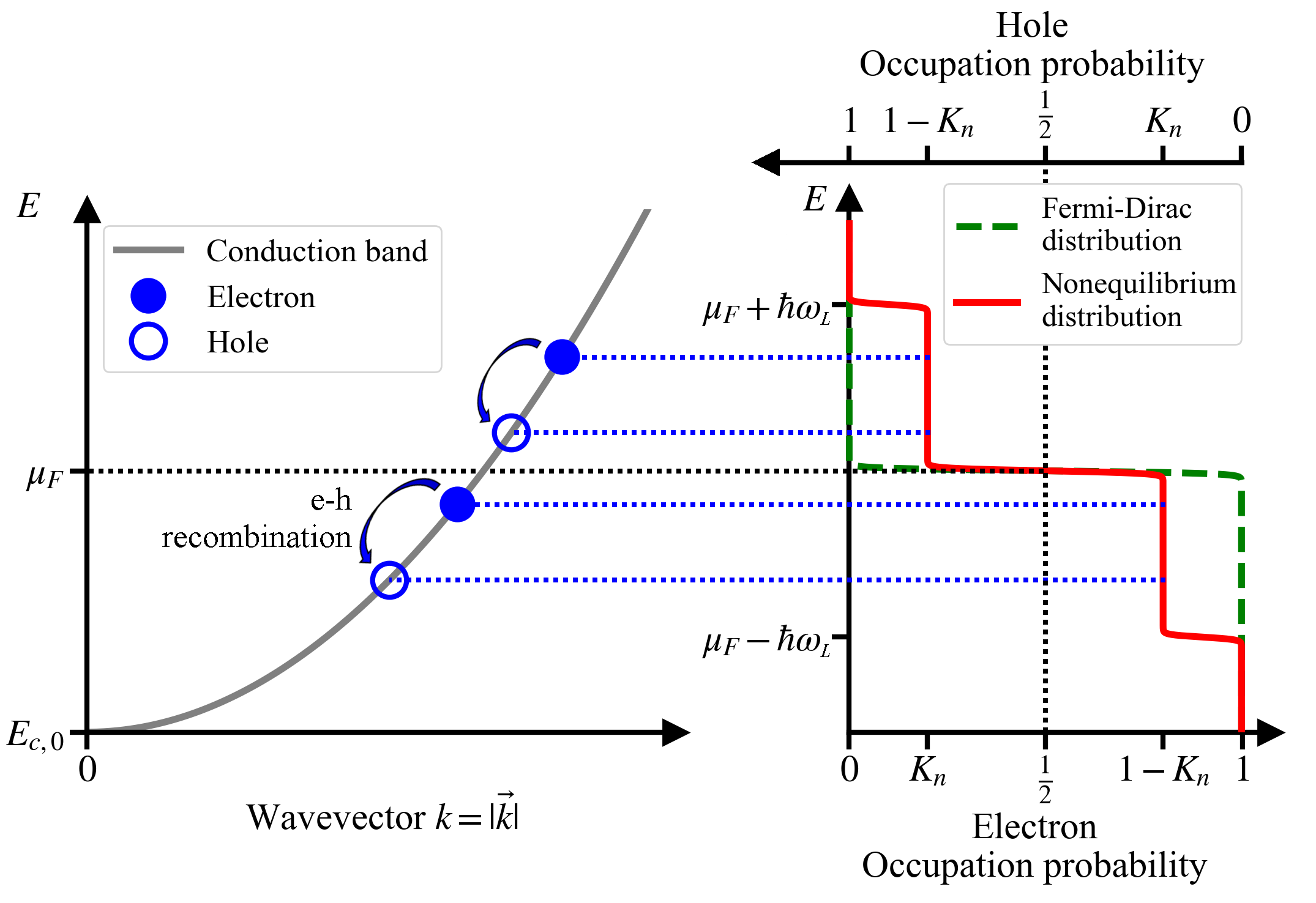}
    \caption{\textit{Nonequilibrium electron-hole recombination (left) and nonequilibrium electron distribution under CW pumping (right)}. In both panels, the Fermi level is depicted by the horizontal, black dashed line at energy $\mu_F$. 
    Right panel: low energy electrons are promoted to states over the Fermi level by absorption of photons with frequency $\omega_L$ from the pumping beam. Accumulation of these hot electrons and their corresponding hot holes is limited by relaxation toward equilibrium, mainly due to electron-electron interaction. In the stationary regime, the competition between excitation and relaxation yields a steady population of hot electrons and holes (red line) with respective occupation probability $K_n$ and $1-K_n$. 
    Left panel: scheme of two electron-hole (e-h) recombinations events, one above the Fermi level and one below, that are enabled by the nonequilibrium distribution of the electrons. The corresponding state occupation probabilities are indicated by the blue dotted lines. 
    Note that an exaggerated value of $K_n \sim 0.2$ has been used to simplify the illustration, when realistic values are $K_n \sim I_L/(10^{11} \text{W.cm}^{-2})$ with $I_L$ the pumping intensity at sample. See Methods for the analytical form of the nonequilibrium distribution and for the interpretation of $K_n$.
    }
    \label{fig:e_h_recombination}
\end{figure}

To start, let us note that CW pumping induces a stationary nonequilibrium electron distribution, whose shape is schematically depicted on the right side of Fig. \ref{fig:e_h_recombination}. It is seen that the nonequilibrium pattern consists in the creation of hot electron and hot hole populations, respectively above and below the Fermi level, with occupation probabilities orders of magnitudes higher compared to equilibrium (though its actual value is exaggerated on the figure for the sake of clarity). Interestingly, an analytical expression of this distribution has been derived in Refs. \citenum{dubi2019hot,sivan2021theory}, using the Boltzmann equation in the relaxation time approximation (see Methods). In Section D. of the Supporting Information, we use this expression to perform the integral $\Theta^{intra}$ in Eq. (\ref{eq:def_teta_gen}). Inserting the result into Eq. (\ref{eq:kirchhoff_generalized}), the emitted power under continuous laser illumination at frequency $\omega_{L}$ can be cast into the form:

\begin{equation}\label{eq:kirchhoff_generalized_pompage_continu}
     \frac{dP_e}{d\Omega_r}(\omega,\boldsymbol{u}_r,p)
     = \sigma_{abs}^{intra} (\omega,-\boldsymbol{u}_r,p^*) \frac{\omega^2}{8\pi^3 c^2} \frac{ 2 (\hbar\omega - \hbar \omega_{L})}{ \exp \big(\frac{\hbar\omega - \hbar\omega_{L}}{k_B T_e} \big) -1}  \frac{\hbar \omega_{L}}{\hbar \omega}  K_n^{eff},
\end{equation}

where $k_B$ is the Boltzmann constant, $T_e$ is the conduction electron temperature, that can be defined rigorously due to the weakness of the nonequilibrium effects (see Methods and Ref. \citenum{dubi2019hot}), and where the dimensionless parameter $K_n^{eff}$ is proportional to the rate of absorbed photons:

\begin{equation}\label{eq:knudsen_eff}
    K_n^{eff} \propto \frac{ \sigma_{abs}^{intra}(\omega_{L},\boldsymbol{u}_{L},p_{L}) I_{L} }{ \hbar \omega_{L} },
\end{equation}

where $\boldsymbol{u}_{L}$, $p_{L}$ and $I_L$ are, respectively,  the direction, the polarization and the intensity at sample of the pumping beam. The full expression of $K_n^{eff}$, given in Methods, shows that it can be identified as an effective Knudsen number and that it does not depend significantly on the emission frequency and electronic temperature.

We can now compare Kirchhoff's law Eq. (\ref{eq:kirchhoff_generalized_pompage_continu}) to the empirical guess Eq. (\ref{eq:spectre_empirique}). Remarkably, we observe that the three terms in the empirical guess are recovered and unambiguously identified in Kirchhoff's law. First, Kirchhoff's law states that the localized plasmon resonance lineshape is given by the product of the absorption cross section by the vacuum mode density, that is $S_{LPR}(\omega) = \sigma_{abs}^{intra,\,eq} (\omega) \times\omega^2/8\pi^3 c^2$. This is an important result, as the shape of the absorption cross section usually differs from the one of the easily measurable scattering cross section, especially for large particles when the electrostatic approximation fails or for complex structures involving several resonances. In particular, we note that our finding is corroborated by Ref. \citenum{link2015} where the PL spectra of Au sphere dimers were shown to resemble more closely the absorption spectra than the scattering spectra. Moreover, let us note that a resonance of the absorption cross section is blueshifted by the $\omega^2$ dependence of the vacuum mode density. This may explain the reproducible blueshift observed, e.g., in Ref. \citenum{shen2012}.

\begin{figure}[htbp]
    \centering
    \includegraphics[width=0.7\columnwidth]{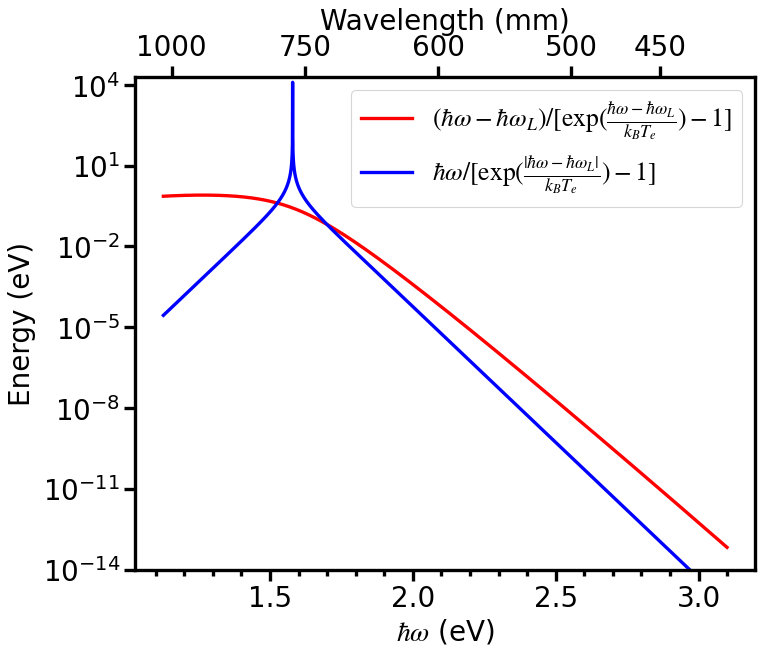}
    \caption{Comparison between the function $F(\hbar \omega-\hbar \omega_L,T_e)/2$ (red line) and a Bose-Einstein distribution with enforced positive frequency $|\omega - \omega_L|$ (blue line). Parameters: $T_e=$ 293 K, $\hbar \omega_L=$ 1.58 eV (i.e. $\lambda_L=$ 785 nm).
    }
    \label{fig:BE_vs_BE_avec_hw_L}
\end{figure}

Second, Kirchhoff's law states that $F(\hbar \omega-\hbar \omega_L,T_e) = 2(\hbar \omega-\hbar \omega_L)/ [ \exp \big(\frac{\hbar\omega - \hbar\omega_{L}}{k_B T_e} \big) -1]$, as previously identified in Ref. \citenum{sivan2021theory}. Interestingly, there is a Bose-Einstein function in $F(\hbar \omega-\hbar \omega_L)$. In the anti-Stokes part of the spectrum, this function enables to recover the exponential scaling $dP_e/d\Omega_r(\omega) \propto \exp \big(-\hbar\omega/k_B T_e \big)$ observed in experiments and used in the anti-Stokes thermometry technique \cite{cahill2014,cahill2016,carattino2018gold,jones2018photothermal,link2019anti,sheldon2019,barella2020,baffou2021}. By contrast, close to $\hbar \omega = \hbar \omega_L$, the shape of the Bose-Einstein distribution is deeply modified by the factor $\hbar\omega - \hbar\omega_{L}$, that turns the divergence of the distribution into a smooth concave curve, as shown on Fig. \ref{fig:BE_vs_BE_avec_hw_L}. This behavior is in agreement with experiments, see, e.g., Ref. \citenum{wen2017spectral} in which the spectrum around $\hbar \omega_L$ is finely probed using an ultra-narrow bandwidth notch filter. 

Further, we stress that the Bose-Einstein function does not originates from a distribution of phonons. On the contrary, it only stems from the electronic distributions product in Eq. (\ref{eq:def_teta_gen}), as shown in Section D. of the Supporting Information. This is the reason why the Bose-Einstein distribution is ruled by the temperature of the conduction electrons. This means that the temperature measured through anti-Stokes thermometry is the electronic temperature, that may differ from the targeted lattice temperature. 

Third, the linear scaling of the Stokes spectrum with the pump intensity is recovered in Kirchhoff's law through the term $K_n^{eff}$ (see Eq. (\ref{eq:knudsen_eff})). In our theory, this term arises from the scheme depicted on Fig. \ref{fig:e_h_recombination}: plasmons and/or photons are absorbed by the metal, that creates "hot" electrons and "hot" holes populations, as shown on the nonequilibrium distribution on the right panel of the figure. Above the Fermi level, a hot electron then recombines with a hole, that emits a plasmon or a photon. The reverse process involving a hot hole is also possible under the Fermi level.  Hence, the overall scheme is based on a one-photon absorption process.

In summary, the considerations above show that Kirchhoff's law Eq. (\ref{eq:kirchhoff_generalized_pompage_continu}) enables to recover the empirical guess Eq. (\ref{eq:spectre_empirique}) and to solve most of the open questions related to this empirical form.

\subsubsection{Quantitative comparisons between theory and experiments}

\begin{figure*}[htbp]
    \centering
    \includegraphics[width=1.0\columnwidth]{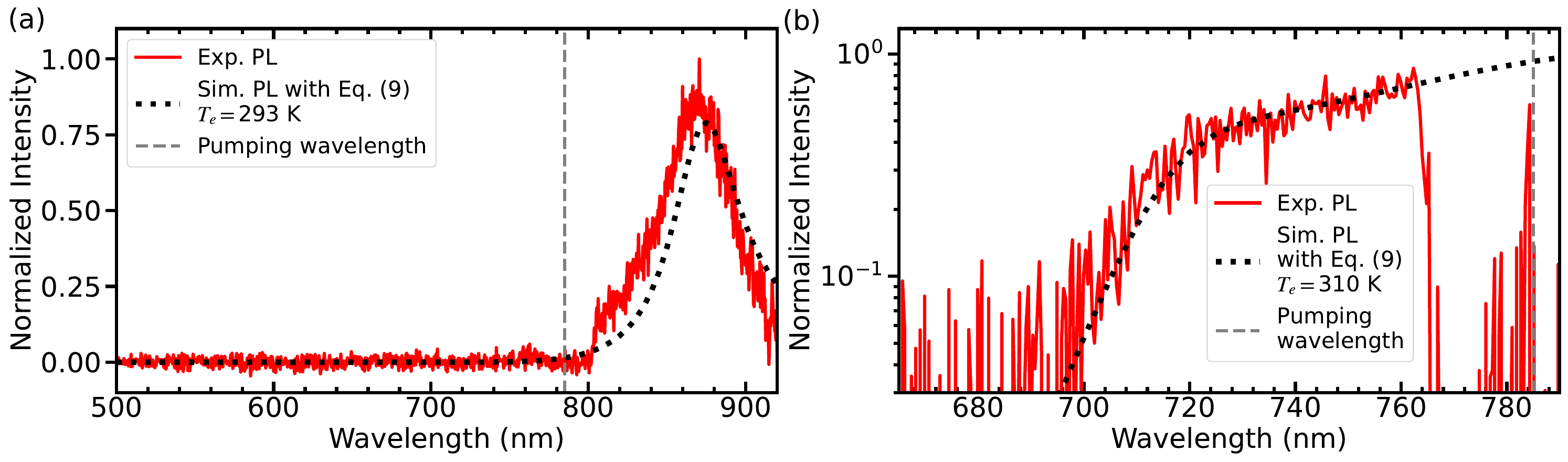}

    \includegraphics[width=1.0\columnwidth]{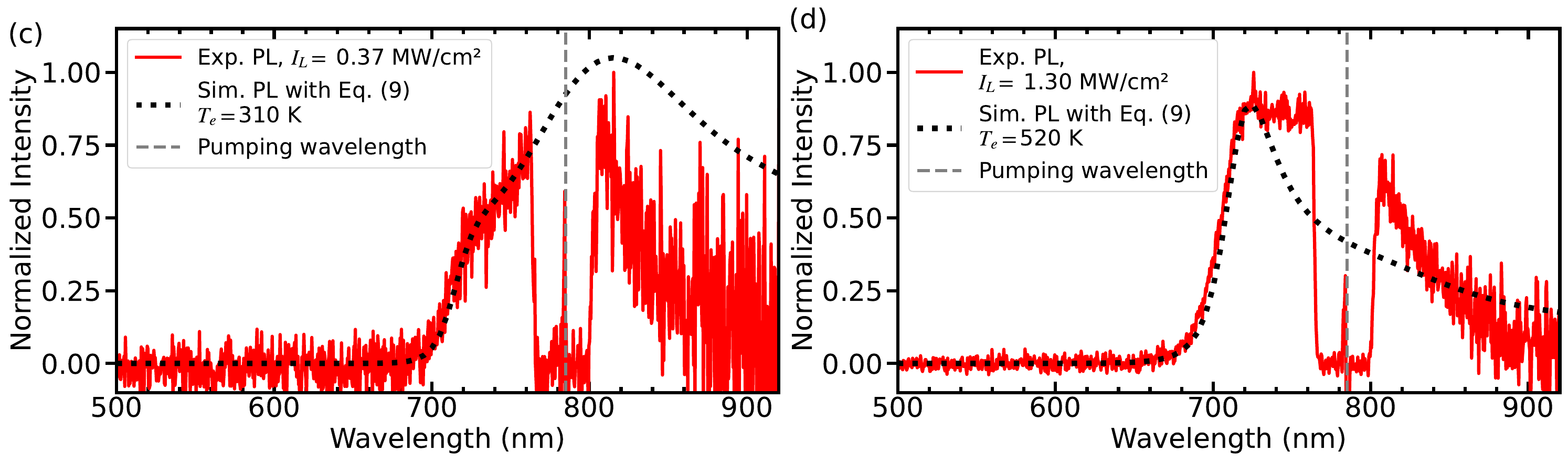}

    \includegraphics[width=1.0\columnwidth]{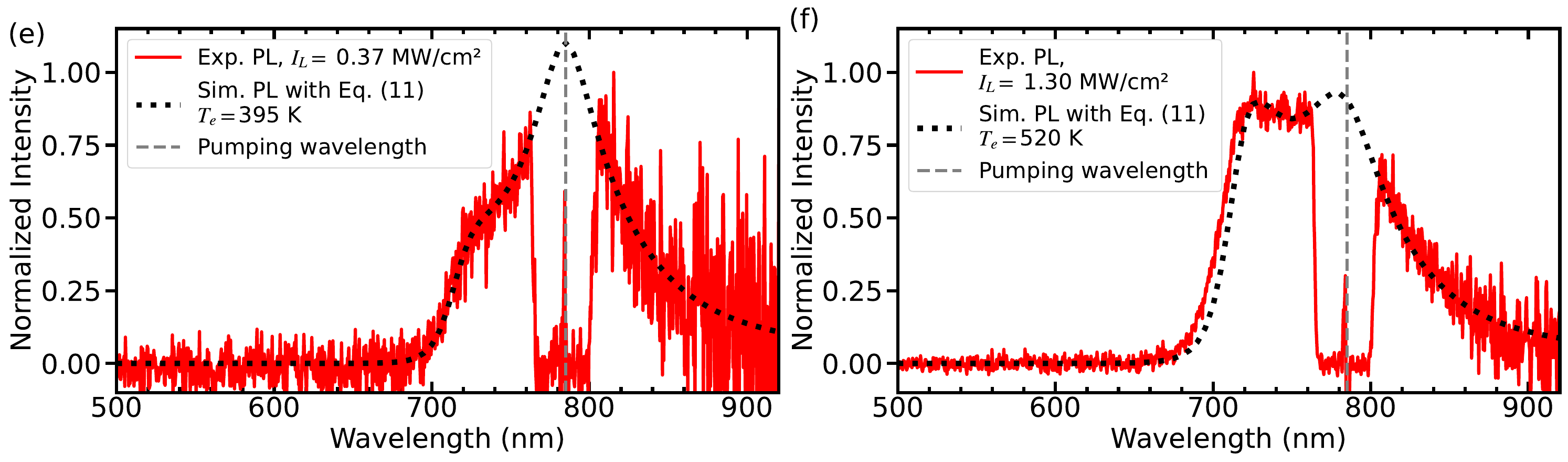}
    \caption{Comparisons between experimental photoluminescence spectra (PL) of gold nanorods (red lines) and simulations with Kirchhoff's law (black dotted lines), under CW pumping at 785 nm. Kirchhoff's law Eq. (\ref{eq:kirchhoff_generalized_pompage_continu}) is used in panels (a-d), while Eq. (\ref{eq:kirchhoff_generalized_CW_pheno}) is used in panels (e-f).
    Panel (a): nanorod with longitudinal plasmon resonance at 871 nm (measured in dark-field scattering) and dimensions 18.2 nm $\times$ 85 nm (measured with SEM). Reprinted with permission from Ref. \citenum{link2018}.
    Panels (b-f): nanorod with longitudinal plasmon resonance at 734 nm and dimensions 25 nm $\times$ 79 nm, illuminated with an intensity at sample $I_L=$ 0.37 MW.cm$^{-2}$ for panels (b-e), and $I_L=$ 1.30 MW.cm$^{-2}$ for panels (d) and (f). The electronic temperature is the only fit parameter and is indicated on the panels. Panel (b) shows same data as panels (c) and (e), but focuses only on the anti-Stokes part of the spectrum and is plotted on semilogarithmic scale.
    Reprinted with permission from Ref. \citenum{link2019anti}.
    Details on the fitting procedure are provided in Methods.
    }
    \label{fig:comparaison_exp_sim}
\end{figure*}

We now turn to a quantitative comparison of the theory with experiments. First, we show on Fig. \ref{fig:comparaison_exp_sim} (a-d) comparisons between gold nanorods emission spectra measured experimentally (red lines) and simulated with Kirchhoff's law (black dotted lines). The position of the pumping wavelength (785 nm) with respect to the longitudinal plasmon resonance enables to focus either on the Stokes spectrum (panel (a)) or on the exponentially decaying part of the anti-Stokes spectrum (panel (b)). In both panels, it is seen that the agreement between theory and experimental data is satisfying. Next, panels (c) and (d) show emission spectra spanning both the Stokes and anti-Stokes sides of the spectrum. Also, these spectra are obtained from the same gold nanorod for two different pumping intensities. It is seen that it induces substantial spectral changes. Adjusting the electronic temperature $T_e$, these spectral changes are well reproduced by the simulations at short wavelengths. Interestingly, it suggests that these changes originate from an increase of the electronic temperature due to the higher pumping intensity. However, it is also seen in both panels that an agreement between theory and experiment cannot be found on the whole spectrum. This points out a limitation of Kirchhoff's law in the form of Eq. (\ref{eq:kirchhoff_generalized_pompage_continu}) that will be addressed in the following subsection. 

Second, let us stress that the theory presented in this work enables for the first time to calculate quantitatively the PL quantum yield. For a gold nanorod pumped continuously at 785 nm and with longitudinal plasmon resonance around 860 nm, corresponding to the experimental conditions of Fig. \ref{fig:comparaison_exp_sim} (a), we calculate in Methods a theoretical quantum yield $\eta_{sim} \sim 1.3 \times 10^{-7}$. Experimentally, the corresponding quantum yield has been estimated to be $\eta_{exp}\sim 1.5 \times 10^{-6}$ in Ref. \citenum{link2018}. While these values do not show a perfect agreement, they remain relatively close to each other. Hence, this shows that light emission through intraband transitions has a high quantum yield and is likely to be a dominant process in PL from metals. This also calls for further refinements of the theory in order to reach a better agreement.

\subsubsection{Phenomenological Kirchhoff's law}

In the previous section, we observed that Kirchhoff's law Eq. (\ref{eq:kirchhoff_generalized_pompage_continu}) reproduces the emission spectrum for short wavelengths but fails for larger wavelengths. Here, we introduce a phenomenological correction to Eq. (\ref{eq:kirchhoff_generalized_pompage_continu}) that enables to solve this issue.

Let us consider here the phenomenological generalization of Kirchhoff's law in the CW pumping regime:
\begin{equation}\label{eq:kirchhoff_generalized_CW_pheno}
     \frac{dP_e}{d\Omega_r}(\omega,\boldsymbol{u}_r,p)
     = \sigma_{abs}^{intra} (\omega,-\boldsymbol{u}_r,p^*) \frac{\omega^2}{8\pi^3 c^2} \times C \frac{ \arctan( \frac{ \hbar\omega - \hbar \omega_{L} }{ k_B T_e })}{ \exp \big(\frac{\hbar\omega - \hbar\omega_{L}}{k_B T_e} \big) -1},
\end{equation}

where $C$ is a constant. We show on Fig. \ref{fig:comparaison_exp_sim} (e),(f) a comparison between this model and the experimental data. The agreement is now satisfying on the whole spectrum in both cases. Remarkably, these fits are achieved using the same simulated absorption cross section, only moderately increasing the electronic temperature $T_e$ as the pumping intensity is raised. We provide further support to the validity of Eq. (\ref{eq:kirchhoff_generalized_CW_pheno}) in Section E. of the Supporting Information. In addition, although our derivation of Eq. (\ref{eq:kirchhoff_generalized_pompage_continu}) and then  Eq. (\ref{eq:kirchhoff_generalized_CW_pheno})  has been restricted to intraband transitions, we have used it to compare with PL spectra obtained using pumping above the interband threshold, so that interband transitions are expected to play a dominant role. It is seen on Figs. S.2 and S.3 of the Supporting Information that the theory is in good agreement with the experimental measurements. This suggests that
a theory for the emission from interband transitions can
be cast into a formula resembling to Eq. (\ref{eq:kirchhoff_generalized_CW_pheno}).

In summary, the comparisons presented in this section suggest that Eq. (\ref{eq:kirchhoff_generalized_CW_pheno}) is a powerful tool with only one free parameter ($T_e$) to predict the emission spectrum of metallic structures pumped in the CW regime. In addition, let us note that Eq. (\ref{eq:kirchhoff_generalized_CW_pheno}) has the same overall form as the generalized Kirchhoff's law Eq. (\ref{eq:kirchhoff_generalized}). Interestingly, this could mean that the missing ingredients to explain the origin of the $arctan$ function in Eq. (\ref{eq:kirchhoff_generalized_CW_pheno}) may be related to the nonequilibrium electronic function $f$ rather than to the approximations underlying Kirchhoff's law Eq. (\ref{eq:kirchhoff_generalized}). This calls for more detailed investigations of the non-equilibrium CW electronic distribution.

\section{Conclusion}

In summary, we have introduced  a model of photoluminescence by metallic structures  based on a statistical physics approach circumventing the calculation of microscopic quantities.

We first derived a nonequilibrium fluctuation relation Eq. (\ref{eq:densite_spectrale_puissance_main}) which relates the current density correlation function with the nonequilibrium electronic distribution. This relation is instrumental in connecting the electron-hole recombination picture of light emission with the Maxwell point of view based on radiation by time-dependent currents. Equipped with this fluctuation relation, we derived the general relation Eq. (\ref{eq:pui_ray_he}) between the power radiated and the nonequilibrium electronic distribution. This relation shows that the absorption cross section can be used to capture the spectral and directional shaping of the emitted light. The full calculation requires a sum over all the transitions including the corresponding populations and matrix elements.

To bypass this cumbersome microscopic approach, we derived in Eq. (\ref{eq:fluctuation_dissipation_theorem_generalized_main}) a  nonequilibrium fluctuation-dissipation relation for intraband transitions. It enabled to establish in Eq. (\ref{eq:kirchhoff_generalized}) a closed-form expression of the emitted power. This form requires only the knowledge of the nonequilibrium macroscopic absorption cross section and the electronic distribution. 

When applied to continuous wave pumping, this theoretical approach recovers the main features of an ansatz used in the literature to analyze the experiments. It enables to resolve several open questions. It also gives a quantitative model of many different published experiments.

This approach has the structure of a generalized Kirchhoff law. We have found that while it is has been derived for intraband transitions, it also enables to reproduce experimental data attributed to interband transitions. This suggests that our derivation of the fluctuation relation and Kirchhoff's law could be extended to interband transitions. This  is left for future work.  Another natural extension of this work is photoluminescence induced by short pulses with high intensities.  In this case, the strong time dependence of the absorptivity due to band filling effects should be included in the generalized Kirchhoff law.  We hope that this  analytical model of photoluminescence will prove  useful to optimize the different applications already demonstrated.

\section{Methods}

\subsection{Nonequilibrium electron distribution under CW pumping and Knudsen number}

We plotted the shape of the nonequilibrium electron distribution under CW pumping on Fig. \ref{fig:e_h_recombination} (right panel, red line). To this end, we used the analytical expression of the conduction electrons distribution under continuous wave pumping $f_{cw}$ that has been derived in Refs. \citenum{dubi2019hot,sivan2021theory}, solving the steady-state Boltzmann equation in the relaxation time approximation for the electron-electron collision rate:
\begin{equation}\label{eq:f_CW}
\begin{split}
    f_{cw}(E,T_e) =  f_{FD}(E,T_e)  + K_n(E,T_e)
     \times [ f_{FD}(E - \hbar \omega_{L},T_e) + f_{FD}(E + \hbar \omega_{L},T_e) - 2f_{FD}(E,T_e) ],
\end{split}
\end{equation}
with the shorthand:
\begin{equation}\label{eq:def_knudsen_app}
   K_n(E,T_e) = \frac{ 4 I_{L} \sigma_{abs}(\hbar \omega_{L},p_{L}) \tau_{ee}(E,T_e) }{ [\hbar \omega_{L}]^2 \rho_c(E)},
\end{equation}

where $E$ stands for the electron energy, $T_e$ the electronic temperature, $f_{FD}(E,T_e)= 1/[ \exp( \frac{E - \mu_F}{k_B T_e} + 1]$ is the Fermi-Dirac distribution with $\mu_F$ the metal Fermi level and $k_B$ the Boltzmann constant, $I_L$, $\omega_L$ and $p_L$ are respectively the intensity, the frequency and the polarization state of the pump beam, $\tau_{ee}$ is the relaxation time associated to electron-electron collisions and $\rho_c$ is the density of electronic states in the conduction band. In transport theory, the dimensionless parameter $K_n$ is called Knudsen number, and quantifies the competition between equilibrium and out-of-equilibrium driving mechanisms. In particular, $K_n$ is the ratio of two characteristic times: (i) $\tau_{ee}(E)$ which is the lifetime of an electron with energy $E$ before it relaxes due to electron-electron collisions, and (ii) $[\hbar \omega_{L}]^2 \rho_c(E)/4 I_{L} \sigma_{abs}(\hbar \omega_{L},p_{L})$ which is the time interval between two consecutive photon absorption events promoting an electron on a final state with energy $E$. Further, it was shown in Refs. \citenum{sivan2019assistance,sivan2021theory} that $K_n \sim I_L/(10^{11} \text{W.cm}^{-2})$, so that $K_n\ll1$ in experimental conditions. Hence, the nonequilibrium correction to the equilibrium distribution is weak and perturbative, so that the electronic temperature $T_e$ can still be rigorously defined \cite{dubi2019hot}. In addition, let us note that electron-phonon collisions are neglected in Eq. (\ref{eq:f_CW}), which is valid provided $|E - \mu_F| \gg k_B T_e$ \cite{sivan2021theory}. Also note that in Fig. \ref{fig:e_h_recombination} we neglected the dependence of $K_n$ on energy $E$ and used an exaggerated value $K_n \sim 0.2$ to simplify the illustration. Finally, we used in Eq. (\ref{eq:knudsen_eff}) an effective Knudsen number, that is related to the above Knudsen number as (see Section D. of the Supporting Information):
\begin{equation}\label{eq:def_Knudsen_eff}
    K_n^{eff} = K_n(E=\mu_F,T_e)\times \bigg[ \frac{\pi k_B T_e}{\hbar \omega_{L}} \bigg]^2 = \frac{4 I_{L} \sigma_{abs}(\hbar \omega_{L},p_L)}{W_{ee}  \rho_c(\mu_F) [\hbar \omega_{L}]^4},
\end{equation}

where $W_{ee}$ is a characteristic electron-electron scattering constant (see Section D. of the Supporting Information). The right hand side of this equation further shows that $K_n^{eff}$ is independent on the electron temperature $T_e$.

\subsection{Numerical procedure for simulations with Kirchhoff's law}

\begin{figure}[htbp]
    \centering
    \includegraphics[width= 0.6\columnwidth]{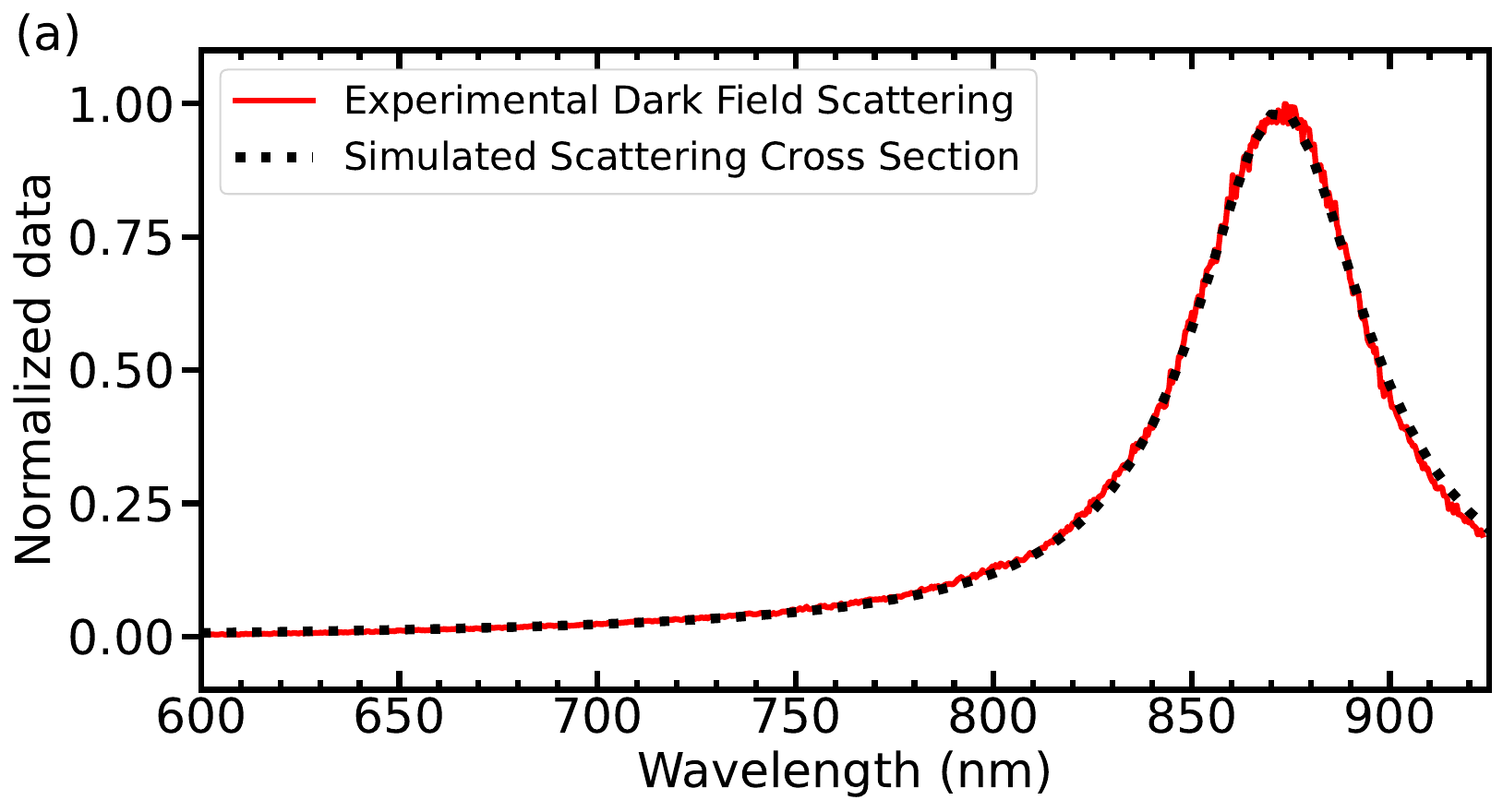}

    \includegraphics[width=0.6\columnwidth]{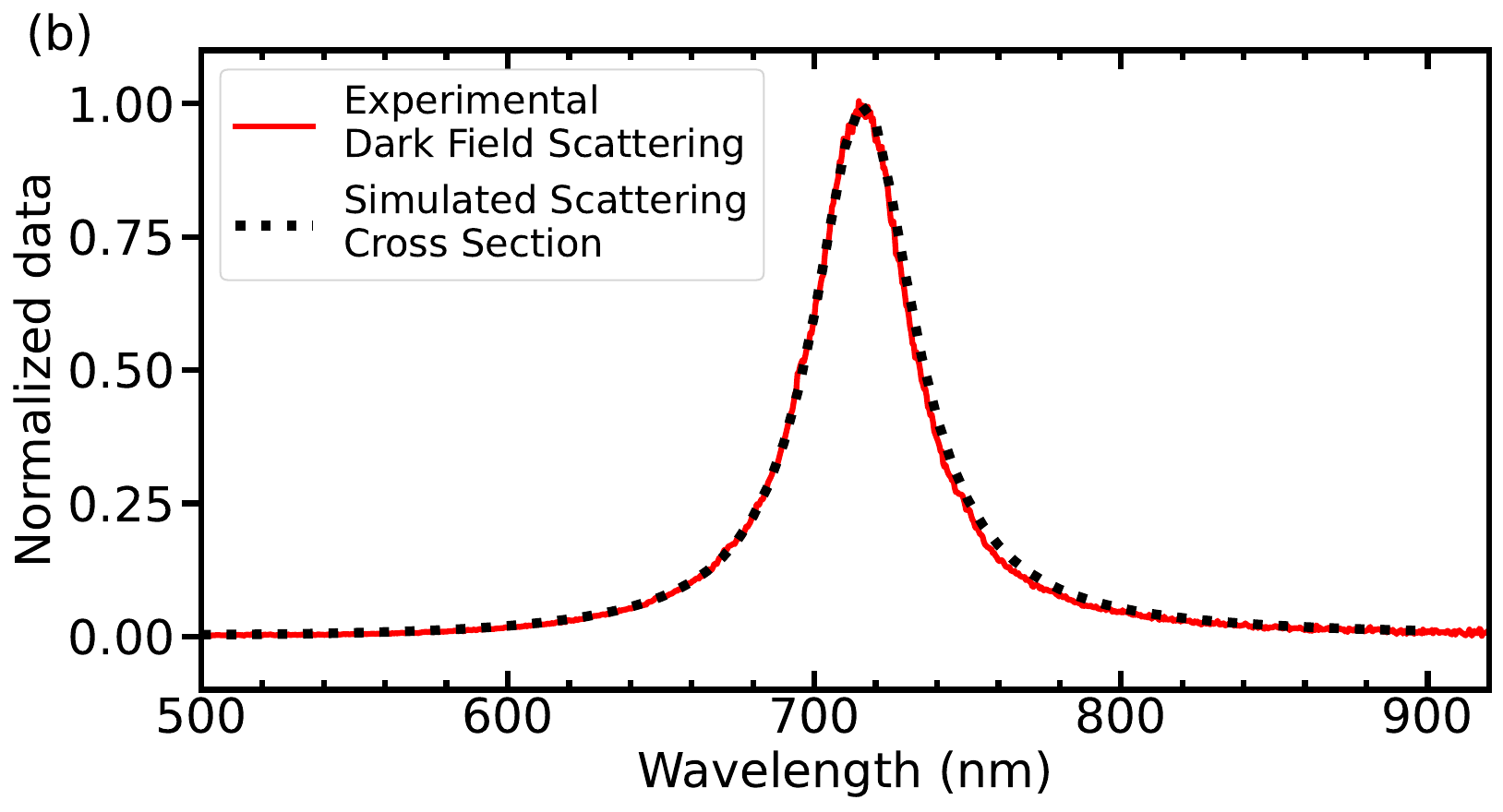}

    \caption{Comparison between the experimental dark-field scattering spectra (red lines) and the simulated scattering cross sections (dark dashed lines) of the gold nanorods studied in Fig. \ref{fig:comparaison_exp_sim} in the main text. Panel (a): nanorod with longitudinal plasmon resonance at 871 nm (Ref. \citenum{link2018}). Dimensions measured with SEM are 18.2 nm $\times$ 85 nm. Adjusted dimensions 18.1 nm $\times$ 85.7 nm are used in the simulation, as well as an absorption enhancement factor $C_{abs}=$ 1.2.
    Panel (b): nanorod with longitudinal plasmon resonance at 734 nm (Ref. \citenum{link2019anti}). Dimensions measured with SEM are 25 nm $\times$ 79 nm. Adjusted dimensions 25 nm $\times$ 78.3 nm are used in the simulation, as well as an absorption enhancement factor $C_{abs}=$ 1.0.
    A Quartz substrate with refractive index 1.4536 has been taken into account.
    }
    \label{fig:DFS_exp_sim}
\end{figure}

We first simulate the scattering cross section corresponding to the metallic structures under study, using the structure dimensions measured experimentally by scanning electron microscopy (SEM). All simulations are carried out with a home-build Maxwell solver based on the Finite Element Method. When possible, we compare the scattering cross section to the experimental dark-field scattering spectra, and adjust the structure dimensions in order to improve their matching. As shown in Fig. \ref{fig:DFS_exp_sim}, a nearly perfect agreement can be found within a sub-nanometer variation of the dimensions for the nanorods studied in Fig. \ref{fig:comparaison_exp_sim} of the main text. We note that our simulations account for the presence of a substrate, so that the experimental conditions are accurately reproduced. In addition, in the simulations illumination is modeled by a plane wave at normal incidence with respect to the substrate. While experimental platforms may use a condenser so that the beam impinges on the nanoparticle with some angle, Fig. \ref{fig:DFS_exp_sim} shows that it does not introduce any flaw. We eventually note that the (normalized) dark-field scattering spectrum and the scattering cross section are defined a bit differently (since the total light collected in the former is reduced compared to the latter due to a limited objective aperture in experiments), but this also does not introduce discrepancies. Finally, we used the values of Ref. \citenum{johnson1972optical} for the gold permittivity. In order to improve the fit quality, the imaginary part of the gold permittivity has been multiplied by a dimensionless factor $C_{abs} > 1$ in some simulations. Indeed, the presence of defects and edges in the narrow nanorods modeled in this work may enhance the absorption processes in the metal, and hence the imaginary part of the permittivity \cite{benisty2022introduction}.

\begin{figure}[htbp]
    \centering
    \includegraphics[width= 0.7\columnwidth]{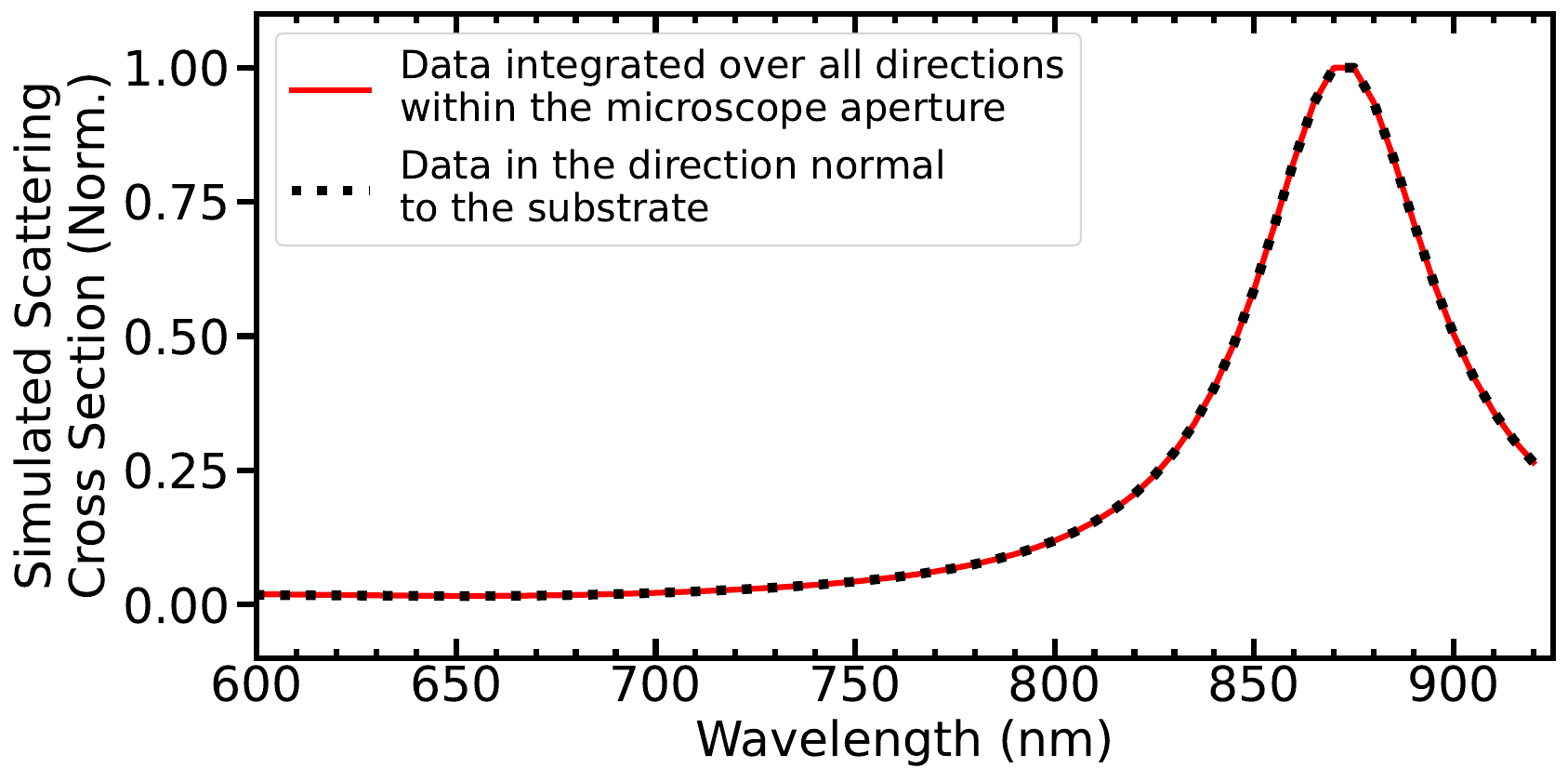}

    \caption{Numerical calculation of the absorption cross section normalized to maximum. The red line shows a calculation taking into account the full numerical aperture of the microscope objective used to collect the emitted light in the experiments (N.A.= 0.8). The dark dotted line shows calculation only at normal incidence with respect to the substrate. The metallic structure used in the calculations correspond to the one whose scattering cross section is shown on Fig. \ref{fig:DFS_exp_sim} (a) and PL is shown on Fig. \ref{fig:comparaison_exp_sim} (a).
    }
    \label{fig:abs_NA_vs_normal_incidence}
\end{figure}

Once the above adjustments are made, the absorption cross section (ACS) of the metallic structure is computed with the same Maxwell solver. On Fig. \ref{fig:abs_NA_vs_normal_incidence}, we show an example of comparison between the normalized ACS calculated at normal incidence with respect to the substrate (dark dotted line) and the normalized ACS calculated taking into account the numerical aperture of the microscope objective used to collect the emitted light in the experiments (red line). No difference can be seen between both quantities, so that the former has been used in the simulations presented in this work (except to calculate the quantum yield in the next section where the full measured signal is needed). We further repeat the calculation of the ACS for orthogonal polarizations. Nevertheless, let us note that absorption in the polarization perpendicular to the nanorods investigated here is negligible. Finally, we stress that only the absorption due to intraband recombinations is to be considered in Fig. \ref{fig:comparaison_exp_sim} of the main text and Fig. S.1 of the Supporting Information. We note that this condition is fulfilled since the contribution from interband recombinations to the imaginary part of the permittivity of gold and silver is negligible in the wavelength ranges studied here.

\subsection{Calculation of the quantum yield in the CW pumping regime with Kirchhoff's law Eq. (\ref{eq:kirchhoff_generalized_pompage_continu})}\label{app:quantum_yield_kirchhoff}

Here, we use Kirchhoff's law Eq. (\ref{eq:kirchhoff_generalized_pompage_continu}) to calculate the PL quantum yield of the gold nanorod whose PL is shown on Fig. \ref{fig:comparaison_exp_sim} (a) of the main text. The theoretical quantum yield $\eta_{sim}$ is evaluated from the power spectrum $dP/d\Omega_r(\omega,\boldsymbol{u}_r,p)$ as:

\begin{equation}
    \eta_{sim}= \frac{ \sum_p \int_{\omega_{min}}^{\omega_{max}} d\omega \int_{\text{N.A.}} d\Omega_r(\boldsymbol{u}_r) \frac{ dP }{ d\Omega_r } (\omega,\boldsymbol{u}_r,p)  }{ \sigma_{abs}(\omega_L,\boldsymbol{u}_L, p_L) I_L },
\end{equation}

where $\omega_{min/max}$ indicates the minimal and maximal frequencies over which the PL signal is integrated, N.A. indicates that the integration runs over the solid angle within the Numerical Aperture of the microscope collecting the PL signal and $\omega_L$, $\boldsymbol{u}_L$, $p_L$, $I_L$ are respectively the frequency, the direction, the polarization and the intensity at sample of the pumping laser. 

In the CW regime, inserting Eq. (\ref{eq:kirchhoff_generalized_pompage_continu}) and Eq. (\ref{eq:def_Knudsen_eff}) into the above equation yields:
\begin{equation}
\begin{split}
    \eta_{sim}= \frac{ 4 }{ W_{ee}  \rho_c(\mu_F)[\hbar \omega_{L}]^4 } \int_{\omega_{min}}^{\omega_{max}}  d\omega & \Bigg[ \sum_p \int_{0}^{\theta_{\text{N.A.}}} sin(\theta) d\theta \int_0^{2\pi}  d\phi\,\, \sigma_{abs}^{intra} (\omega,\theta,\phi,p^*) \Bigg] \\
     & \times \frac{\omega^2}{8\pi^3 c^2} \frac{ 2 (\hbar\omega - \hbar \omega_{L})}{ \exp \big(\frac{\hbar\omega - \hbar\omega_{L}}{k_B T_e} \big) -1}  \frac{\hbar \omega_{L}}{\hbar \omega},
\end{split}
\end{equation}

where we used the spherical coordinate angles $\theta$ and $\phi$ to parameterize the emission angle ($\theta$ being defined from the normal direction with respect to the substrate). We then use the following numerical values that are representative of the experimental conditions of Ref. \citenum{link2018}: (i) $W_{ee}=$ 0.01135 eV$^{-2}$.fs$^{-1}$, \footnote{This value enables the best fit with Eq. (S.39) of the function used in Ref. \citenum{sivan2021theory}: $\tau_{ee}(E)= e^{\frac{1}{a+b(E-\mu_F)}}$ with $a=$ 0.08585 and $b=$ 0.1278 eV$^{-1}$ in the range 1.24-1.57 eV. } (ii) $\rho_c(\mu_F)= [V/2\pi^2]\times [2m_{e}/\hbar^2]^{3/2} \times \sqrt{\mu_F}$ in the parabolic approximation for the conduction band, with $V= \pi\times 9.1^2\times 85$ nm$^3$ \cite{link2018}, $m_{e}$ the electron mass and $\mu_F=$ 5.5 eV \cite{ashcroft2022solid}, (iii) $\hbar \omega_L=$ 1.58 eV, $\omega_{min}=$ 1.24 eV and $\omega_{max}=$ 1.57 eV \cite{link2018}, (iv) $\theta_{\text{N.A.}} = \arcsin(0.8)$ \cite{link2018}, (v) $T_e$= 293 K. Using these values and calculating numerically the absorption cross section according to the procedure described in the last section, we obtain the theoretical estimate of the quantum yield:

\begin{equation}
    \eta_{sim} \sim 1.3\times 10^{-7}.
\end{equation}

\begin{acknowledgement}

We thanks M. Besbes for assistance with finite element simulations, as well as Y. Cai, J. Tauzin and S. Link for sharing with us their experimental data. This work was supported by Agence Nationale de la Recherche (GYN project ANR-17-CE24-0046).

\end{acknowledgement}

\begin{suppinfo}
\end{suppinfo}

\setcounter{equation}{0}
\setcounter{figure}{0}
\setcounter{table}{0}
\renewcommand{\theequation}{S.\arabic{equation}}
\renewcommand{\thefigure}{S.\arabic{figure}}
\renewcommand{\thetable}{S.\arabic{table}}

\section{A. Statistical physics model of light emission by a reciprocal medium}\label{app:link_power_psd}

In this section, we introduce the key elements of a statistical physics model of light emission by a reciprocal medium. In particular, we link the electromagnetic power radiated by a metallic structure to the cross spectral density of its electronic current density. We also show that accounting for Lorentz reciprocity enables to reformulate the radiation problem into a practical form using the absorption cross section.

\subsection{A.1. Linking the emitted power spectrum to the cross spectral density of the electronic current density}

Let us consider an emitting metallic structure, with a basis centered somewhere in the structure. We focus on an observation point $\boldsymbol{r}$ in the far field. The corresponding emission direction is defined by the unit vector $\boldsymbol{u}_r= \boldsymbol{r}/|\boldsymbol{r}|$. The power $dP_e^{}(t,\boldsymbol{r},p)$ emitted by the metallic structure at time $t$, in a polarization state $p$ and measured by a detector with surface $dA$ placed at the point $\boldsymbol{r}$ is given by the Poynting vector \cite{loudon2000quantum}:
\begin{equation}\label{eq:puissance_mesuree}
    dP_{e}^{}(t,\boldsymbol{r},p) = 2\epsilon_0 c \langle \hat{ {\mathcal{E}}}^{(-)}(t,\boldsymbol{r},p) \hat{{\mathcal{E}}}^{(+)}(t,\boldsymbol{r},p) \rangle\, dA,
\end{equation}

where $\hat{{\mathcal{E}}}^{(+/-)}(t,\boldsymbol{r},p) = \boldsymbol{e}^{(p)} \cdot \hat{\boldsymbol{\mathcal{E}}}^{(+/-)}(t,\boldsymbol{r})$ is the amplitude of the electric field operator along the unit vector $\boldsymbol{e}^{(p)}$ defining the polarization state $p$, with $\hat{\boldsymbol{\mathcal{E}}}^{(+)}(t,\boldsymbol{r}) = \int_{0}^{+\infty} \hat{\boldsymbol{\mathcal{E}}}(\omega,\boldsymbol{r}) e^{-i \omega t} d\omega/2\pi\,\,$ the positive frequency part of the electric field operator $\hat{\boldsymbol{ \mathcal{E} }}(t,\boldsymbol{r})$, and $\hat{\boldsymbol{\mathcal{E}}}^{(-)}$ the negative frequency part, that satisfies $\hat{\boldsymbol{\mathcal{E}}}^{(-)} = \hat{\boldsymbol{\mathcal{E}}}^{(+)\,\dagger}$. In this work, $\langle ... \rangle$ stands for the canonical ensemble average. We note that the order $(-),(+)$ of the electric field operators corresponds to normal ordering, that stems from the absorption process occurring in the detector.

First, let us remark that the electric field operators appearing in the right hand side of Eq. (\ref{eq:puissance_mesuree}) are connected to the current density operator $\hat{\boldsymbol{j} }(\omega)$ associated to a homogeneous metal through:
\begin{equation}\label{eq:green_tensor}
    \hat{ {\mathcal{E}}}(\omega,\boldsymbol{r},p) = i \mu_0 \omega \int_V d^3 \boldsymbol{r}' e^{(p)}_{x} G_{xy}(\omega,\boldsymbol{r},\boldsymbol{r}\,') \hat{j}_{y} (\omega)
\end{equation}

where $\mu_0$ is the vacuum permeability, $V$ is the metal volume and $G_{xy}$ is the component of the Green tensor $\boldsymbol{G}$ associated to the vector components $x$ and $y$. Throughout this work, we use the Einstein notation so that there is a sum over repeated indices. 

Second, we remind that the cross spectral density $W_{j^{\dagger}j} (\omega)$ of a homogeneous and stationary system is defined from the correlation of its current density \cite{goodman2015statistical}:
\begin{equation}\label{eq:correlation_courrant_psd_1}
     \langle \hat{j}_x^{\dagger}(\omega) \hat{j}_{y}(\omega') \rangle = 2\pi\delta(\omega - \omega') \frac{1}{V} 
     W_{j^{\dagger}_x j_y}^{} (\omega),
\end{equation}

where $\delta$ is the Dirac function. 

Hence, inserting Eq. (\ref{eq:green_tensor}) into the Fourier transform of Eq. (\ref{eq:puissance_mesuree}) and using Eq. (\ref{eq:correlation_courrant_psd_1}), the power spectrum can be cast into the form:
\begin{equation}\label{eq:pui_em_psd_x_y}
    \frac{dP_e}{d\Omega_r}  (\omega,\boldsymbol{r},p) =  \frac{|\boldsymbol{r}|^2 \omega^2}{\pi \epsilon_0 c^3 }  \int_V d^3 \boldsymbol{r}' e^{(p)}_{u} G_{ux}(\omega,\boldsymbol{r}, \boldsymbol{r}')e^{(p)}_{v} G_{vy}(\omega,\boldsymbol{r}, \boldsymbol{r}') W_{j^{\dagger}_x j_y}(\omega),
\end{equation}

where we have used $d\Omega_r = dA/|\boldsymbol{r}|^2$ to denote the solid angle subtended by the surface $dA$.

Finally, noting that the permittivity of common metals is isotropic, the cross spectral density tensor reduces to $W_{j^{\dagger}_x j_y}= \delta_{x,y}W_{j^{\dagger}j}$, so that Eq. (\ref{eq:pui_em_psd_x_y}) can be simplified into:
\begin{equation}\label{eq:pui_em_psd}
    \frac{dP_e}{d\Omega_r} (\omega,\boldsymbol{r},p) =   \frac{|\boldsymbol{r}|^2 \omega^2}{\pi \epsilon_0 c^3 } W_{j^{\dagger}j}^{}(\omega) \times \sum_x \int_V d^3 \boldsymbol{r}' |e^{(p)}_{u} G_{ux}(\omega,\boldsymbol{r}, \boldsymbol{r}')|^2.
\end{equation}

\subsection{A.2. Role of Lorentz reciprocity}

The emitted power in Eq. (1) of the main text is expressed as a function of the absorption cross section of the metallic body. We now show how to relate Eq. (\ref{eq:pui_em_psd}) to this quantity \cite{Greffet2018}. To start, we recall the definition of the absorption cross section of an isotropic and homogeneous metallic body of volume $V$:
\begin{equation}\label{eq:def_sec_abs}
    \sigma_{abs}(\omega,-\boldsymbol{u}_r, p) = \frac{\omega}{c} \text{Im}[\epsilon(\omega)] \int_{V_{}} d\boldsymbol{r}' \frac{|\boldsymbol{{\mathcal{E}}}(\boldsymbol{r}',\omega,-\boldsymbol{u}_r,p)|^2}{|\mathcal{E}_{L} (\omega,-\boldsymbol{u}_r,p)|^2},
\end{equation}

where $\epsilon$ is the permittivity of the metal, ${{\mathcal{E}}}_{L}(\omega,-\boldsymbol{u}_r, p)$ is the electric field amplitude corresponding to an incident plane wave and $\boldsymbol{{\mathcal{E}}}(\boldsymbol{r}', \omega, -\boldsymbol{u}_r, p)$ is the electric field induced by this plane wave at a position $\boldsymbol{r}'$ in the metal. It will be convenient to assume that the incident plane wave is emitted by an electric dipole with frequency $\omega$ and polarization $p$ placed at the point $\boldsymbol{r}$ in the far field of the metallic body. This emission point defines the plane wave direction $-\boldsymbol{u}_r= -\boldsymbol{r}/|\boldsymbol{r}|$.

In this context, the vector component $u$ of $\boldsymbol{ \mathcal{E}}(\boldsymbol{r}', \omega, -\boldsymbol{u}_r, p)$ can be related to $\mathcal{E}_{L}(\omega, -\boldsymbol{u}_r, p)$ using the Green tensor:
\begin{equation}\label{eq:relation_geen_Einc}
    \mathcal{E}_{u}(\boldsymbol{r}', \omega, -\boldsymbol{u}_r, p) = 4 \pi |\boldsymbol{r}| e^{ - \frac{i \omega |\boldsymbol{r}|}{c} } G_{ux} (\omega,\boldsymbol{r}', \boldsymbol{r}) e^{(p)}_{x} \mathcal{E}_{L}(\omega, -\boldsymbol{u}_r, p).
\end{equation}

Hence, inserting Eq. (\ref{eq:relation_geen_Einc}) into Eq. (\ref{eq:def_sec_abs}) yields:
\begin{equation}\label{eq:def_sec_abs_green}
    \sigma_{abs} (\omega,-\boldsymbol{u}_r, p) = \frac{\omega}{c} \text{Im}[\epsilon(\omega)] (4\pi |\boldsymbol{r}|)^2 \times \sum_u \int_{V} d^3 \boldsymbol{r}' |G_{ux} (\omega,\boldsymbol{r}', \boldsymbol{r}) e^{(p)}_{x}|^2.
\end{equation}

Since the metal has a scalar permittivity, Lorentz reciprocity imposes \cite{Collin60}:
\begin{equation}\label{eq:reciprocite_lorentz}
    G_{xu}(\omega,\boldsymbol{r},\boldsymbol{r}\,') = G_{ux}(\omega,\boldsymbol{r}\,',\boldsymbol{r}).
\end{equation}

Using Eq. (\ref{eq:reciprocite_lorentz}), the absorption cross section Eq. (\ref{eq:def_sec_abs_green}) can be inserted into Eq. (\ref{eq:pui_em_psd}), which yields the desired connection between the emitted power, the cross spectral density of $\hat{\boldsymbol{j}}$ and the absorption cross section:
\begin{equation}\label{eq:pui_em_psd_abs}
    \frac{dP_e}{d\Omega_r}(\omega,\boldsymbol{u}_r,p) =  \sigma_{abs}(\omega,-\boldsymbol{u}_r, p)  \frac{ \omega^2}{8 \pi^3 c^2 }   \frac{ W_{j^{\dagger}j} (\omega) }{ 2 \omega \epsilon_0 \text{Im}[\epsilon(\omega)] }.
\end{equation}

This last equation corresponds to the Equation (1) reported in the main text.

\section{B. Fluctuation and dissipation of a stationary nonequilibrium electron gas}\label{app:psd_electron_gas}

In this section, we derive nonequilibrium forms of $W_{j^{\dagger}j}$ and $\text{Im}[\epsilon]$. We start by calculating the cross spectral density of the current density $W_{j^{\dagger}j} (\omega)$, that quantifies the system fluctuations. We then derive $\text{Im}[\epsilon]$, through a calculation of the linear response under optical excitation. Finally, we show that when the gas is in thermodynamic equilibrium, $W_{j^{\dagger}j} (\omega)$ can be related to the linear response, yielding the so-called Fluctuation-Dissipation theorem.

\subsection{B.1. Cross spectral density of the current density}\label{app:calculation_psd}

The current density operator of a homogeneous metal at a time $t$ in the Heisenberg picture can be expressed generally as a function of the electron field operator $\hat{\psi}$ \cite{giuliani2005quantum}:
\begin{equation}\label{eq:def_current_density_operator}
    \hat{ \boldsymbol{j} }(t) = \frac{e}{2 m V} \int_V d^3 \boldsymbol{r} \Big[ \hat{\psi}^{\dagger}(\boldsymbol{r},t) \textbf{p}  \hat{\psi}(\boldsymbol{r},t) - \hat{\psi}(\boldsymbol{r},t) \textbf{p} \hat{\psi}^{\dagger}(\boldsymbol{r},t) \Big],
\end{equation}

where $e$ is the electron charge, $m$ is the electron mass, $V$ is the metal volume, $\boldsymbol{r}$ is a position in the metal, $^{\dagger}$ is the Hermitian conjugate and $\textbf{p} = -i\hbar \nabla$ is the momentum operator with $\hbar$ the reduced Planck constant and $\nabla$ the gradient operator.

First, we focus on the electron field operator. In the following, we will assume that the electron gas can be described as a non-interacting gas, and will neglect the electron spin degree of freedom. Accordingly, the electronic states are given by the solutions of the single-electron Schrödinger equation in the metal (with a mean-potential accounting for the lattice). We index these states by $n=1...N$ where $N$ is the total electron number, and note their corresponding energy by $\hbar \omega_n$ and wavefunction by $|\phi_{n} \rangle$. Hence, 
the second-quantization Hamiltonian describing the electron gas writes: 
\begin{equation}
    \hat{\mathcal{H}} = \sum_{n=1}^{N} \hbar \omega_n \hat{c}_{n}^{\dagger} \hat{c}_{n}^{},
\end{equation}

where $\hat{c}_{n}^{\dagger}$, $\hat{c}_{n}$ are respectively the electron creation and annihilation operators of the state $n$, that obey the Fermionic commutation relations $\{ \hat{c}_{n}^{\dagger},\hat{c}_{n'}^{\dagger} \} =0$; $\{ \hat{c}_{n}^{},\hat{c}_{n'}^{} \} =0$ and $\{ \hat{c}_{n}^{\dagger},\hat{c}_{n'}^{} \} = \delta_{n,n'}$, where $\{\hat{a}, \hat{b}  \}= \hat{a}\hat{b} + \hat{b} \hat{a}$. In this context, the electron field operator in the Heisenberg picture is defined as \cite{giuliani2005quantum}:
\begin{equation}\label{eq:def_electron_field_operator}
    \hat{\psi}_{ }(\boldsymbol{r},t) = \sum_{n}  \phi_{n}(\boldsymbol{r}) e^{-i\omega_n t} \hat{c}_{n},
\end{equation}

where $\phi_{n}(\boldsymbol{r}) = \langle \boldsymbol{r} | \phi_{n} \rangle$. Inserting the electron field operator definition Eq. (\ref{eq:def_electron_field_operator}) into Eq. (\ref{eq:def_current_density_operator}) and carrying out a time Fourier transform, we obtain the current density operator in the frequency domain:
\begin{equation}\label{eq:operateur_densite_courant_frequency_domain}
   \hat{ j }_{x}(\omega) = \frac{e}{m V}  \sum_{n,n'}  2 \pi \delta( \omega - [\omega_{n'} - \omega_{n}] ) p^{x}_{n,n'}  \hat{c}_{n}^{\dagger} \hat{c}_{n'},
\end{equation}

where $x$ indicates a vector component in a given basis and where we introduced the shorthand $ p^{x}_{n,n'}  = \langle \phi_n| p^{x} | \phi_{n'} \rangle$. 

Then inserting Eq. (\ref{eq:operateur_densite_courant_frequency_domain}) into Eq. (\ref{eq:correlation_courrant_psd_1}) defined in the last section, we can calculate  the correlation of the current densities. To this end, the main difficulty to overcome is the evaluation of the 4-operators correlations $\langle  \hat{c}_{n'}^{\dagger} \hat{c}_{n} \hat{c}_{m}^{\dagger} \hat{c}_{m'}\rangle$. Yet, let us remind that for a non-interacting electron gas, {the electronic states can be described as Fock states, that we note by $|\{ n_l \}\rangle= |n_1...n_l... n_N \rangle$ where $n_l$ denotes the number of particles in the state $l$. The explicit form of the average correlation is $\langle  \hat{c}_{n'}^{\dagger} \hat{c}_{n} \hat{c}_{m}^{\dagger} \hat{c}_{m'}\rangle = \sum_{ \{ n_l \} } P_{\{ n_l \}} \langle \{ n_l \} | \hat{c}_{n'}^{\dagger} \hat{c}_{n} \hat{c}_{m}^{\dagger} \hat{c}_{m'} | \{ n_l \} \rangle$, where $P_{\{ n_l \}}$ is the canonical probability that the system is in the Fock state $|\{ n_l \} \rangle$. The elements in the sum can be evaluated given that $\langle \{ n_l \} | \hat{c}_{n}^{\dagger} | \{ n_l \} \rangle = \langle \{ n_l \} | \hat{c}_{n}^{} | \{ n_l \} \rangle = 0 $ and $\langle \{ n_l \} | \hat{c}_{n}^{\dagger} \hat{c}_{n'}^{} | \{ n_l \} \rangle = \delta_{n,n'} f_{n}$, where $f_{n}$ is the mean occupation number of the electronic state $n$. These considerations enables to simplify the above correlations:
\begin{equation}\label{eq:4_operator_correlation}
    \langle  \hat{c}_{n'}^{\dagger} \hat{c}_{n} \hat{c}_{m}^{\dagger} \hat{c}_{m'}\rangle= \delta_{n,n'} \delta_{m,m'} \Big[ \delta_{n,m} f_{n} + [1-\delta_{n,m}]f_{n}f_{m} \Big] + \delta_{n',m'} \delta_{n,m} [1-\delta_{n,n'}] f_{n'} [1 - f_{n}].
\end{equation}

Importantly, let us stress that at this point the mean occupation numbers can take any value, that is the $f_{n}$ may differ from the equilibrium Fermi-Dirac distribution. We also note that the first term in the right hand side of Eq. (\ref{eq:4_operator_correlation}) only implies emission at zero frequency so that it is not relevant to the present study. Conversely, the second term in the right hand side of Eq. (\ref{eq:4_operator_correlation}) involves mixing of the $n$ and $m$ electronic states, namely time-dependent correlation between the two current densities. Using this relation finally enables to derive the fluctuation relation reported in Eq.(2) of the main text:
\begin{equation}\label{eq:power_spectral_density}
     W_{j^{\dagger}j}^{} (\omega) = \frac{\epsilon_0 \omega_p^2}{m} \frac{2\pi}{N} \sum_{n,n'=1}^{N} |p^{}_{n,n'}|^2 f_{n'}[1 - f_{n}] \delta( \omega - [\omega_{n'} - \omega_{n}] ),
\end{equation}

where $\epsilon_0$ is the vacuum permittivity and $\omega_p~=~\sqrt{Ne^2/mV\epsilon_0}$ is the plasma frequency.

In summary, we have provided a derivation of the nonequilibrium cross spectral density of $\hat{\boldsymbol{j}}$. We note that a similar expression was used in Ref. \citenum{roques2022framework} in the context of scintillation-induced light emission. To finish, let us discuss further the transition matrix elements $|p^{}_{n,n'}|=|\langle \phi_n| \textbf{p} | \phi_{n'} \rangle|$ that appears in Eq. (\ref{eq:power_spectral_density}). We note that for a bulk, non-interacting electron gas, the electronic wavefunctions reduce to Bloch waves so that $|\langle \phi_n| \textbf{p} | \phi_{n'} \rangle| = 0$ if $n\neq n'$. In reality, these transition matrix elements are obviously non-zero, otherwise the existence of thermal emission would be precluded. Actually, it is well-known that intraband absorption processes are assisted by phonons or electrons. While taking into account these microscopic processes is beyond the scope of the present paper, we note that this issue has been tackled in Ref. \citenum{pottier2021physique} (Chapter 15. Section 3.) for the contribution of static impurities and in Ref. \citenum{loirette2023etude} (Chapter 5. Appendix A.) for the contribution of phonon-assisted transitions.

\subsection{B.2. Linear response}

We now calculate the imaginary part of the permittivity of a nonequilibrium electron gas in the linear response approximation. We start by writing the semiclassical coupling Hamiltonian in the Schrödinger picture between an incident radiation described by the vector potential $\boldsymbol{A}$ and a homogeneous electron gas described by the electron current density $\hat{ \boldsymbol{j}}$:
\begin{equation}
    \hat{\mathcal{H}}_{int}(t) = - \hat{ \boldsymbol{j}}. \boldsymbol{A}(t) V,
\end{equation}

where we have neglected spatial variations of the vector potential inside the metallic structure, in accordance with the homogeneous gas description.


We then treat the Hamiltonian $\hat{\mathcal{H}}_{int}(t)$ as a time-dependent perturbation of the electron gas. In the linear response approximation, the so-called Kubo formula \cite{kubo1957statistical,kubo1957statistical2} expresses elegantly the expectation value of $\hat{\boldsymbol{j}}^{ind}$, the electron current density induced by the perturbation: 
\begin{equation}\label{eq:formule_Kubo}
    \langle \hat{j}^{ind}_{x}(t) \rangle = \langle \hat{j}_{x}(t_0) \rangle  - \frac{iV}{\hbar} \int_{t_0}^{t} dt'\langle [ \hat{j}_{x}(t),  \hat{j}_{y} (t')] \rangle {A}_{y}(t'),
\end{equation}

where it should be noted that $\hat{\boldsymbol{j}}_{ind}(t)$ is written in the interaction picture while $\hat{ \boldsymbol{j}}(t)$ is now written in the Heisenberg picture. 

From Eq. (\ref{eq:formule_Kubo}), it is possible to define the electric susceptibility tensor of the electron gas in the \textit{vector potential-current density} gauge (noted by $\boldsymbol{j}.\boldsymbol{A}$):
\begin{equation}\label{eq:def_tenseur_suseptibilite}
    \chi_{x,y}^{\boldsymbol{j}.\boldsymbol{A}}(t,t') =  - \frac{i V}{\hbar }  \langle [\hat{j}_{x}(t),  \hat{j}_{y} (t')] \rangle H(t-t'),
\end{equation}

where $H$ is the Heaveside step function. Now, resorting again to the assumption that the electrons are non-interacting, the current density operator in the Heisenberg picture can be expressed by inserting Eq. (\ref{eq:def_electron_field_operator}) into Eq. (\ref{eq:def_current_density_operator}). Then inserting the result into Eq. (\ref{eq:def_tenseur_suseptibilite}) and using the relation Eq. (\ref{eq:4_operator_correlation}) to simplify the 4-operator correlation, we obtain $\chi^{\boldsymbol{j}.\boldsymbol{A}}$ into the form:
\begin{equation}\label{eq:tenseur_suseptibilite_temporel}
     \chi_{x,y}^{\boldsymbol{j}.\boldsymbol{A}} (t-t')= \frac{i}{\hbar}\frac{\epsilon_0 \omega_p^2}{m} \frac{1}{N} \sum_{n,n'} p^{x}_{n,n'} p^{y}_{n',n} [f_{n'} - f_{n}] e^{-i (\omega_{n'} - \omega_n)(t-t')} H(t-t').
\end{equation}

Finally, moving to the electric field-dipole moment gauge and noting that Eq. (\ref{eq:tenseur_suseptibilite_temporel}) is stationary in time, the imaginary part of the permittivity tensor (that is the imaginary part of the electric susceptibility tensor) associated to metal writes in the frequency domain:
\begin{equation}\label{eq:tenseur_permittivite_spectral_non_iso}
     \text{Im}[\epsilon_{x,y}](\omega) = \frac{\omega_p^2}{2\hbar m \omega^2} \frac{2\pi}{N} \sum_{n,n'=1}^{N} p^{x}_{n,n'}  p^{y}_{n',n} [f_{n} - f_{n'}] \delta( \omega - [\omega_{n'} - \omega_{n}] ).
\end{equation}

In the assumption that the metal is isotropic, this equation reduces to:
\begin{equation}\label{eq:tenseur_permittivite_spectral}
     \text{Im}[\epsilon_{}^{}](\omega) = \frac{\omega_p^2}{2\hbar m \omega^2} \frac{2\pi}{N} \sum_{n,n'=1}^{N} |p^{}_{n,n'}|^2 [f_{n} - f_{n'}] \delta( \omega - [\omega_{n'} - \omega_{n}] ),
\end{equation}

where it is stressed again that the $f_{n}$ can take any value.

\subsection{B.3. Fluctuation-Dissipation theorem}

To finish, let us focus on the peculiar but common case in which the electron gas has relaxed to thermodynamic equilibrium. The mean occupation numbers $f_{n}$ of the electrons then follow the Fermi-Dirac distribution:
\begin{equation}\label{eq:fermi_dirac_distribution}
    f_n= f_{FD}(\omega_n,T_e,\mu_F) \equiv \frac{1}{\exp \Big( \frac{\hbar \omega_n - \mu_F}{k_B T_e} \Big) + 1},
\end{equation}

where $T_e$ is the temperature of the electron gas and $\mu_F$ its Fermi level. Using Eq. (\ref{eq:fermi_dirac_distribution}), it is straightforward to derive the identity:
\begin{equation}\label{eq:ratio_distributions}
    \frac{ f_{FD}(\hbar \omega_n) - f_{FD}(\hbar \omega_n + \hbar \omega) }{ \exp \Big( \frac{\hbar \omega}{k_B T} \Big) - 1}
    = f_{FD}(\hbar \omega_n + \hbar \omega) [ 1-  f_{FD}(\hbar \omega_n) ].
\end{equation}

The right hand side of Eq. (\ref{eq:ratio_distributions}) can be viewed as the probability of spontaneous emission in a photonic mode with energy $\hbar \omega$ whereas the left hand side is the product of the probability of absorption by the Bose-Einstein distribution. This equation is actually equivalent to the so-called Van Roosbroeck-Shockley relation, which connects the emission and absorption rates in a homogeneous semiconductor medium \cite{van1954photon,loirette2023photon}.

Using Eq. (\ref{eq:ratio_distributions}), a connection can now be made between the cross spectral density Eq. (\ref{eq:power_spectral_density}) and the linear response Eq. (\ref{eq:tenseur_permittivite_spectral}) of the electron gas at thermodynamic equilibrium, that is called the Fluctuation-Dissipation theorem \cite{callen1951irreversibility,kubo1966fluctuation}:
\begin{equation}\label{eq:FDT_Wjj_appendix}
    W_{j^{\dagger}j}^{eq} (\omega,T_e) = 2 \omega \epsilon_0 \text{Im}[\epsilon_{eq}(\omega,T_e)] \frac{\hbar \omega}{ \exp \big( \frac{\hbar \omega}{k_B T_e} \big) - 1},
\end{equation}

which corresponds to equation (3) of the main text.

\section{C. Generalized Kirchhoff's law for emission from intraband transitions}\label{app:generalized_kirchhoff_law}

In this section, we detail the approximations and calculation steps used to derive Eq. (7) of the main text. Hence, we focus on the conduction band electronic states and intraband transitions only. We want to calculate the cross spectral density corresponding to these intraband recombination processes, that we note $W_{j^{\dagger}j}^{intra}$. 

We start by assuming that the transition matrix elements of intraband transitions are constant over the electronic states, and equal to their value at Fermi level. Building on the notations of section B., this amounts to write that ${p}_{n,n'}= {p}_{F}$. This assumption enables to cast Eq. (\ref{eq:power_spectral_density}) into:
\begin{equation}\label{eq:power_spectral_density_pF}
     W_{j^{\dagger}j}^{intra} (\omega) = \frac{\epsilon_0 \omega_{p,c}^2}{m} \frac{2\pi }{N_c} |p^{}_{F}|^2 \sum_{n,n'=1}^{N_c} f_{n'}[1 - f_{n}] \delta( \omega - [\omega_{n'} - \omega_{n}] ),
\end{equation}

where $ \omega_{p,c} = \sqrt{N_c e^2/mV\epsilon_0}$ with $N_c$ the number of electrons in the conduction band. We stress that now the sums over the electronic states is restricted to the conduction band. 

We now convert the sum over electronic states into integrals. To this end, we assume that the mean occupation numbers $f_n$ of the conduction band electronic states follow a distribution $f$ that only depends on the state energy. Hence, Eq. (\ref{eq:power_spectral_density_pF}) can be written as:
\begin{equation}\label{eq:power_spectral_density_generalized}
     W_{j^{\dagger}j}^{intra} (\omega) = \frac{\epsilon_0 \omega_{p,c}^2}{m} \frac{2\pi }{N_c} |p^{}_{F}|^2 D_J(\mu_F,\hbar \omega) \int_{E_{c,0}}^{\infty} d E \, f(E + \hbar \omega) [1 - f(E)],
\end{equation}

where $E$ is an electron state energy, $E_{c,0}$ the energy minimum of the conduction band, and $D_J(E,\hbar\omega)$ is the energy distribution of the joint density of states (defined by volume unit):
\begin{equation}
    D_{J}(\omega,E)= V^2 \int_{E_{c,0}}^{\infty} d E' \, \rho_c(E') \rho_c(E)\delta(\hbar \omega - [E' - E]),
\end{equation}

with $\rho_{c}(E)$ the density of states of the conduction band (defined by volume unit). In addition, let us note that in Eq. (\ref{eq:power_spectral_density_generalized}) we assumed $D_{J}(\omega,E)$ to be constant over the electronic states and equal to its value at Fermi level, so that it can be taken out of the integral.

We now focus on the prefactor of the integral in Eq. (\ref{eq:power_spectral_density_generalized}). Performing the same approximations as above on Eq. (\ref{eq:tenseur_permittivite_spectral}), the imaginary part of the metal permittivity corresponding to absorption through intraband processes can be cast into the form:
\begin{equation}\label{eq:tenseur_permittivite_spectral_generalized}
     \text{Im}[\epsilon^{intra}](\omega)= \frac{\omega_{p,c}^2}{2\hbar m \omega^2} \frac{2\pi }{N_c} |p_F|^2 D_J(\mu_F,\omega) \int_{E_{c,0}}^{\infty} d E \, [f(E) - f(E + \hbar \omega)].
\end{equation}

At thermodynamic equilibrium, it is possible to simplify further Eq. (\ref{eq:tenseur_permittivite_spectral_generalized}). Indeed, the distributions $f$ can then be replaced by Fermi-Dirac distributions. Interestingly, we further note that the ambient-temperature thermal energy ($\sim$ 25 meV) is negligible compared to the energy of visible or NIR photons ($\sim$ 1 eV). Therefore, the Fermi-Dirac distribution can be approximated by
a step function, that is $f_{FD}(E) \approx 1 - H(E-\mu_F)$. Inserting this approximation into Eq. (\ref{eq:tenseur_permittivite_spectral_generalized}), we find that at thermodynamic equilibrium $\text{Im}[\epsilon^{intra}]$ reduces to:
\begin{equation}\label{eq:tenseur_permittivite_spectral_generalized_encore_plus}
     \text{Im}[\epsilon^{intra}_{eq}](\omega)=  \frac{\omega_{p,c}^2}{2\hbar m \omega^2} \frac{2\pi }{N_c} |p_F|^2 D_{J}(\mu_F,\hbar \omega) V \hbar \omega.
\end{equation}

Finally, inserting Eq. (\ref{eq:tenseur_permittivite_spectral_generalized_encore_plus}) into Eq. (\ref{eq:power_spectral_density_generalized}) yields a generalization of the Fluctuation-Dissipation theorem, valid for any out-of-equilibrium electronic distribution $f$:
\begin{equation}\label{eq:fluctuation_dissipation_theorem_generalized_app}
     W_{j^{\dagger} j}^{intra} (\omega)= 2\epsilon_0 \omega \text{Im}[\epsilon^{intra}_{eq}(\omega)] \int_{E_{c,0}}^{\infty} d E \, f(E + \hbar \omega) [1 - f(E)],
\end{equation}

that is the equation reported in Eqs. (5),(6) of the main text. 

Now, inserting Eq. (\ref{eq:fluctuation_dissipation_theorem_generalized_app}) into Eq. (\ref{eq:pui_em_psd_abs}) yields:
\begin{equation}\label{eq:kirchhoff_generalized_app}
     \frac{dP_e}{d\Omega_r}(\omega,\boldsymbol{u}_r,p)
     = \sigma_{abs}^{}(\omega,-\boldsymbol{u}_r,p^*) \frac{\omega^2}{8\pi^3 c^2} \frac{\text{Im}[\epsilon^{intra}_{eq}(\omega)]}{\text{Im}[\epsilon^{}(\omega)]} \int f(E + \hbar \omega) [ 1-  f(E) ] dE.
\end{equation}

We finally define the absorption cross section due to intraband transitions only as:
\begin{equation}\label{eq:abs_cross_sec_intra}
     \sigma_{abs}^{intra}(\omega,-\boldsymbol{u}_r,p^*) = \frac{\text{Im}[\epsilon^{intra}(\omega)]}{\text{Im}[\epsilon^{}(\omega)]} \times  \sigma_{abs}^{}(\omega,-\boldsymbol{u}_r,p^*),
\end{equation}

where we note that the imaginary part of the metal permittivity due to intraband transitions $\text{Im}[\epsilon^{intra}(\omega)]$ is not necessarily evaluated at equilibrium here. Inserting Eq. (\ref{eq:abs_cross_sec_intra}) into Eq. (\ref{eq:kirchhoff_generalized_app}), we get:  
\begin{equation}\label{eq:kirchhoff_generalized_app_intra}
     \frac{dP_e}{d\Omega_r}(\omega,\boldsymbol{u}_r,p)
     = \sigma_{abs}^{intra}(\omega,-\boldsymbol{u}_r,p^*) \frac{\omega^2}{8\pi^3 c^2} \frac{\text{Im}[\epsilon^{intra}_{eq}(\omega)]}{\text{Im}[\epsilon^{intra}(\omega)]} \int f(E + \hbar \omega) [ 1-  f(E) ] dE. 
\end{equation}

Finally, assuming that $\text{Im}[\epsilon^{intra}_{eq}(\omega)]/\text{Im}[\epsilon^{intra}(\omega)] \approx 1$, namely that possible nonequilibrium effects are sufficiently weak so that the imaginary part of the metal permittivity stays close to the equilibrium one, we obtain the equation (7) reported in the main text. We note that this last approximation is expected to hold in most pumping conditions, in particular through continuous wave pumping. However, it may be inaccurate for excitation through ultrashort laser pulses with high peak intensity \cite{hou2018absorption}.

\section{D. Generalized Kirchhoff's law for emission from intraband transitions in the CW regime}

In this section, we detail the derivation of the generalized Kirchhoff's law for emission from intraband transitions in the CW pumping regime Eq. (9) from the generalized Kirchhoff's law Eq. (7).

To start, we insert Eqs. (12), (13) of Methods into Eq. (6) of the main text. Keeping only the components to first order in $K_n$, the integral $\Theta^{intra}$ can be cast into the form:
\begin{equation}
\begin{split}
    \Theta^{intra}_{cw}(\omega) = & \int f_{FD}(E + \hbar \omega - \hbar \omega_{L}) [ 1-  f_{FD}(E)] K_n(E + \hbar \omega) dE \\
    & + \int f_{FD}(E + \hbar \omega) [1 - f_{FD}(E + \hbar \omega_{L})] K_n(E) dE. 
\end{split}
\end{equation}

Using the remarkable relation between Fermi-Dirac distributions: 
\begin{equation}
    \frac{ f_{FD}(E+E_1) [ 1 - f_{FD}(E+E_2) ] }{ f_{FD}(E+E_2)  - f_{FD}(E+E_1)} = \frac{1}{ \exp(\frac{E_1 - E_2}{k_B T_e}) - 1},
\end{equation}

we get:
\begin{equation}\label{eq:theta_gen_CW_intermediaire}
\begin{split}
   \Theta^{intra}_{cw}(\omega) = \frac{1}{ \exp \big(\frac{\hbar\omega -\hbar \omega_{L}}{k_B T_e} \big) - 1}  \times \Bigg[  & \int [ f_{FD}(E) - f_{FD}(E + \hbar \omega - \hbar \omega_{L} ) ] K_n(E + \hbar \omega) dE \\
    &  + \int [ f_{FD}(E + \hbar \omega_{L}) - f_{FD}(E + \hbar \omega) ] K_n(E) dE \Bigg].
\end{split}
\end{equation}

As in Section C., we now use the fact that thermal energy $k_B T$ at ambient temperature is negligible compared to $\hbar \omega,\, \hbar \omega_{L},\, \mu_F$. Therefore, we approximate the Fermi-Dirac distribution by a step function $f_{FD}(E) \approx 1 - H(E-\mu_F)$. Hence, Eq. (\ref{eq:theta_gen_CW_intermediaire}) simplifies into:
\begin{equation}\label{eq:theta_gen_CW_intermediaire_2}
   \Theta^{intra}_{cw}(\omega) = \frac{1}{ \exp \big(\frac{\hbar\omega -\hbar \omega_{L}}{k_B T_e} \big) - 1}  \times \Bigg[  \int_{\mu_F + \hbar \omega_L }^{\mu_F + \hbar \omega_{} } K_n(E) dE  + \int_{\mu_F - \hbar \omega}^{\mu_F - \hbar \omega_{L} } K_n(E) dE \Bigg].
\end{equation}

We then need to perform an integral over the Knudsen number. To simplify this integral, we assume that the electronic density of state $\rho_c$ can be evaluated at Fermi level, as in the last section when we derived the generalized Kirchhoff's law. Hence, according to Eq. (13) in Methods, the only quantity that depends on energy in $K_n$ is the relaxation time $\tau_{ee}$. In Landau’s Fermi liquid
Theory \cite{coleman2015book}, this dependence is a Lorentzian:
\begin{equation}\label{eq:tau_ee_fermi_liquid}
    \tau_{ee}(E,T_e) = \frac{1}{W_{ee}[\pi k_B T_e]^2}  \frac{1}{ 1 + \Big[\frac{E - \mu_F}{\pi k_B T_e} \Big]^2 },
\end{equation}

where $W_{ee}$ is a characteristic electron-electron scattering constant.

Using Eq. (\ref{eq:tau_ee_fermi_liquid}) into Eq. (\ref{eq:theta_gen_CW_intermediaire_2}), one finds an analytic result involving $\arctan$ functions. Using the remarkable identity $\arctan(x) - \arctan(y) = \arctan(\frac{x - y }{1-xy})$ with $x,y > 0$ eventually enables to cast $\Theta^{intra}(\omega)$ under CW pumping into the form:

\begin{equation}
   \Theta^{intra}_{cw}(\omega) = \frac{2 ( \hbar \omega - \hbar \omega_{L}) }{ \exp \big( \frac{\hbar\omega - \hbar \omega_{L}}{k_B T_e} \big) - 1} \frac{\hbar \omega_{L}}{\hbar \omega_{}} K_n^{eff},
\end{equation}

where we defined the effective Knudsen number also given in Methods:
\begin{equation}
    K_n^{eff} = K_n(E=\mu_F,T_e)\times \bigg[ \frac{\pi k_B T_e}{\hbar \omega_{L}} \bigg]^2 = \frac{4 I_{L} \sigma_{abs}(\hbar \omega_{L},p_L)}{W_{ee}  \rho_c(\mu_F) [\hbar \omega_{L}]^4}
\end{equation}

that is independent on the electron temperature $T_e$.

\section{E. Phenomenological Kirchhoff's law: additional simulation-experiment comparisons}\label{app:pheno_kichhoff_law}

In this appendix, we provide additional evidence that the phenomenological Kirchhoff's law Eq. (11) is an accurate tool to model PL from metals in the CW pumping regime. We reproduce data obtained with silver and data measured with gold pumped at 532 nm that exhibit PL in the 500-600 nm range.

\begin{figure}[htbp]
    \centering
    \includegraphics[width=0.7\columnwidth]{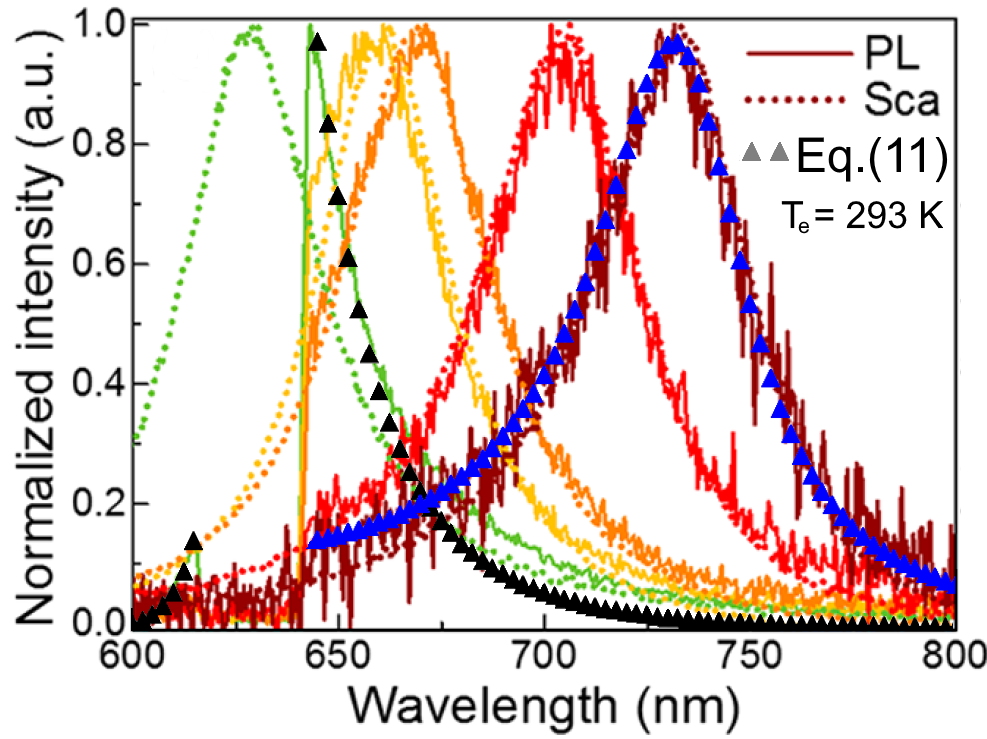}
    \caption{Comparison between experimental \cite{ren2016} photoluminescence spectra (PL) from silver nanorods  of various lengths (plain colored lines) and simulations with the phenomenological Kirchhoff's law Eq. (11) (black and blue triangles). The colored dotted lines correspond to experimental dark-field scattering spectra and evidence the longitudinal plasmon resonance of the nanorods. The width of the nanorods was measured to 23 nm \cite{ren2016}. 
    All nanorods are pumped at 633 nm.
    The theory-experiment comparison lies on the nanorods with respectively lower (exp: green line; sim: black triangles) and higher (exp: brown line; sim: blue triangles) resonance wavelength. Their length is estimated to 72 nm and 92 nm respectively through adjustments of the dark-field scattering spectra with the scattering cross section (not shown). These adjustments also required to use $C_{abs}=1.8$ and $C_{abs}=3.5$ respectively. Figure adapted from Ref. \citenum{ren2016}.
    Details on the fitting procedure are provided in Methods.
    }
    \label{fig:comparaison_exp_sim_silver}
\end{figure}

First, we show on Fig. \ref{fig:comparaison_exp_sim_silver} experimental PL spectra of silver nanorods with different length under CW pumping at 633 nm (plain colored lines) \cite{ren2016}. This material is interesting as it does not feature interband transitions in the visible range, which enables to study intraband PL at shorter excitation wavelengths than we did previously for gold. For readability reasons, we only show comparisons with simulations (colored triangles) for the nanorods with respectively lower (exp: green line, sim: black triangles) and higher (exp: brown line, sim: blue triangles) resonance wavelength. Again, the agreement between theory and experiments is very good for both nanorods. In particular, let us notice that the green curve features a small anti-Stokes emission around 620 nm that is nicely reproduced by the simulation. Hence, this provides support that calculations with Eq. (11) are robust whatever type of emitting metal.

\begin{figure}[htbp]
    \centering
    \includegraphics[width=0.7\columnwidth]{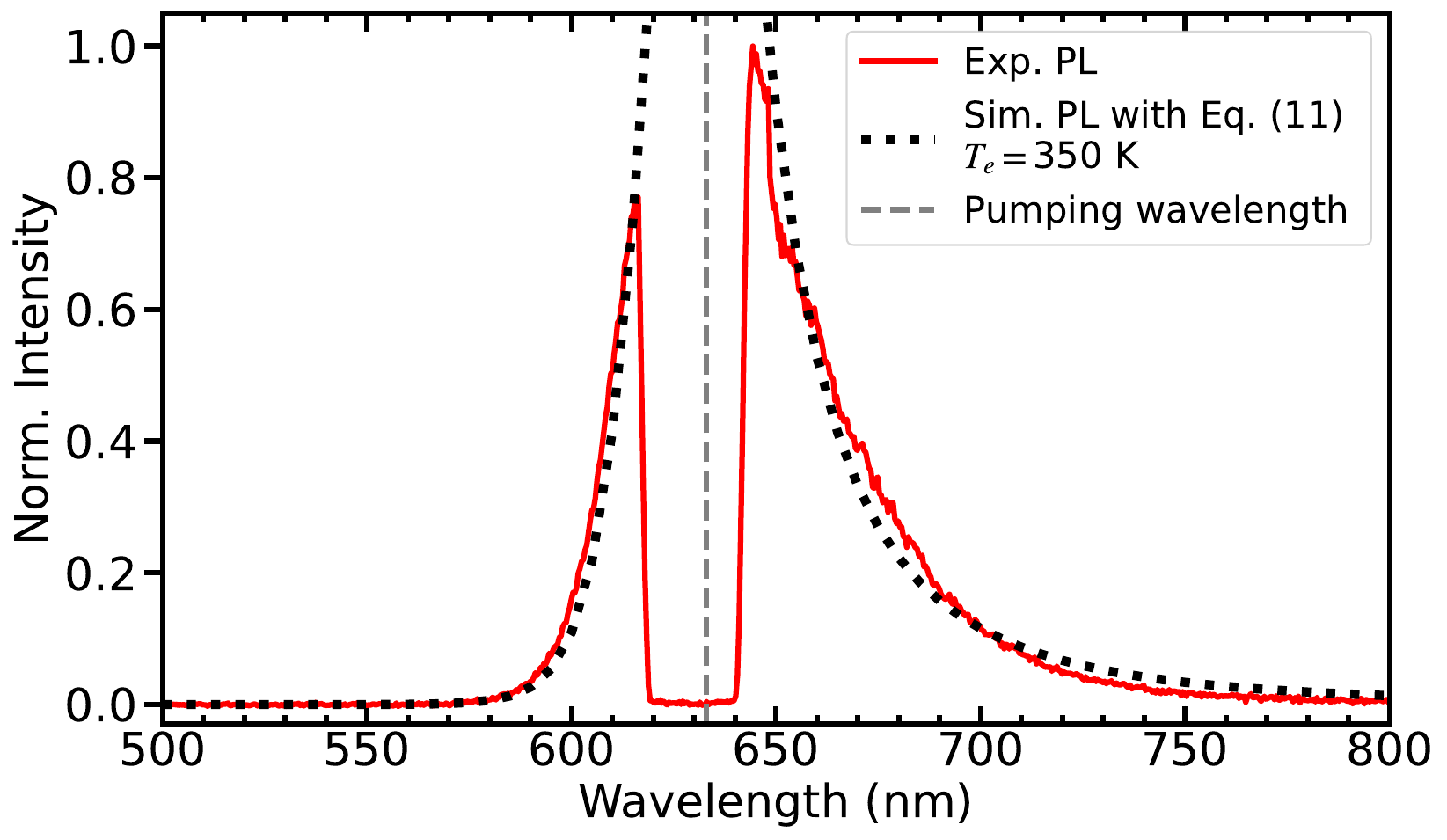}

    \caption{Comparison between the experimental \cite{link2019anti} photoluminescence spectrum (PL) of a gold nanorod (red line) and a simulation with the phenomenological Kirchhoff's law Eq. (11) (black dashed line), under pumping at 633 nm.
    The nanorod has a longitudinal plasmon resonance at 619 nm (measured in dark-field scattering) and dimensions 30 nm $\times$ 63 nm (measured with SEM). Reprinted with permission from Ref. \citenum{link2019anti}. The electron temperature is the only fit parameter and is indicated on the figure.
    Details on the fitting procedure are provided in Methods.
    }
    \label{fig:comparaison_exp_sim_gold_nanorod_633}
\end{figure}

\begin{figure}[htbp]
    \centering
    \includegraphics[width=0.7\columnwidth]{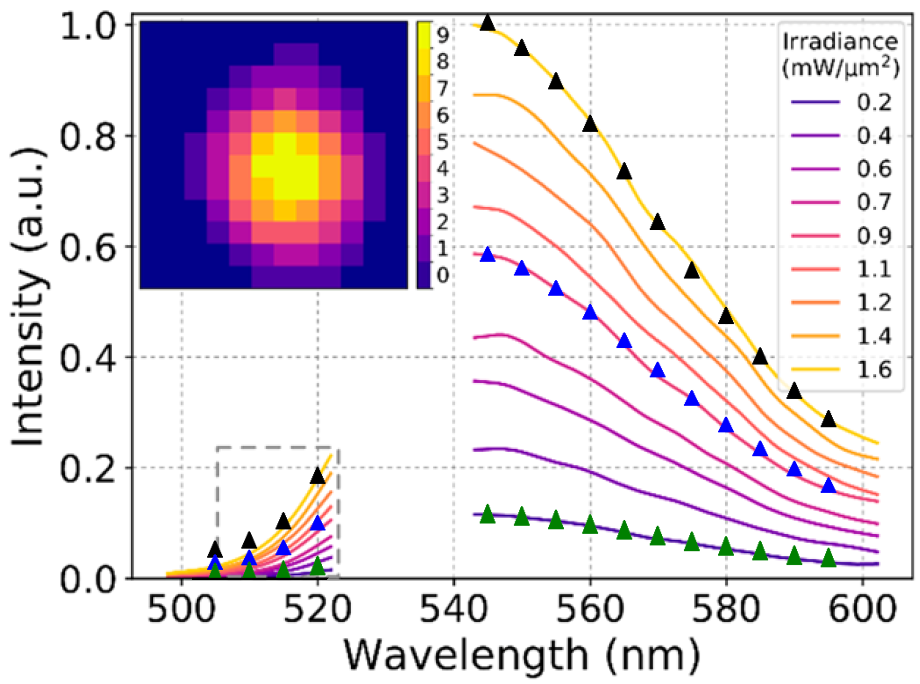}
    \caption{Comparison between the experimental \cite{barella2020} photoluminescence spectra (PL) of a gold nanosphere in water (colored lines) and simulations with the phenomenological Kirchhoff's law Eq. (11) (colored triangles), under pumping at 532 nm and for several illumination intensities.
    The nanosphere has a longitudinal plasmon resonance at 540 nm and diameter 80 nm. 
    The theory-experiment comparison is made for the illumination irradiances 0.2 mW.$\mu$m$^2$ (exp: purple line; sim: green triangles), 0.9 mW.$\mu$m$^2$ (exp: pink line; sim: blue triangles) and 1.6 mW.$\mu$m$^2$ (exp: yellow line; sim: black triangles). The electron temperature is the only fit parameter, and is 293 K for the green and blue triangles and 320 K for the black triangles.
    The absorption cross section of the nanosphere in water has been calculated with a Mie theory solver. The influence of the substrate has been neglected. 
    Figure adapted from Ref. \citenum{barella2020}.
    }
    \label{fig:comparaison_exp_sim_gold_nanosphere_532}
\end{figure}

Second, we show on Figs. \ref{fig:comparaison_exp_sim_gold_nanorod_633} and \ref{fig:comparaison_exp_sim_gold_nanosphere_532} experimental PL spectra for metallic nanostructures pumped in CW over the interband threshold (Gold nanorod pumped at 633 nm in Fig. \ref{fig:comparaison_exp_sim_gold_nanorod_633}; Gold nanosphere pumped at 532 nm in Fig. \ref{fig:comparaison_exp_sim_gold_nanosphere_532}). While Kirchhoff's law Eq. (11) has been developed for CW emission under the interband threshold, we try to fit with it the experimental data, only replacing $\sigma_{abs}^{intra}$ by $\sigma_{abs}^{}$. It is seen on both figures that a quite good agreement can be reached. We stress that interband transitions are expected to provide the leading contribution in this frequency range. Hence, this agreement suggests that the contribution of the interband transitions can also be cast in a form resembling Eq. (11).

\bibliography{achemso}

\end{document}